\documentclass{article}

\usepackage{arxiv}

\usepackage[utf8]{inputenc} 
\usepackage[T1]{fontenc}    
\usepackage{url}            
\usepackage{booktabs}       
\usepackage{amsfonts}       
\usepackage{nicefrac}       
\usepackage{microtype}      
\usepackage{lipsum}		
\usepackage{graphicx}
\usepackage{natbib}
\usepackage{doi}
\usepackage{graphicx}
\usepackage{amsmath}
\usepackage{gensymb}
\usepackage{dcolumn}
\usepackage{bm}
\usepackage{natbib}
\usepackage{subcaption}
\usepackage{hyperref}
\hypersetup{
  colorlinks   = true, 
  urlcolor     = blue, 
  linkcolor    = blue, 
  citecolor   = blue 
}
\usepackage[mathlines]{lineno}

\title{General relativistic limit of $f(R)$ gravity scalaron mass through black hole shadow}

\author{ \href{https://orcid.org/0009-0005-5021-9996}{\includegraphics[scale=0.06]{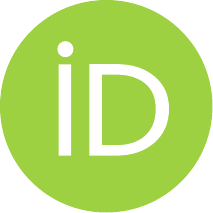}\hspace{1mm}Pranjali Bhattacharjee}\\
	Department of Physics\\
	Gauhati University\\
	Jalukbari, Guwahati-781014, Assam, India \\
	\texttt{pranjali@gauhati.ac.in} \\
	\And
	\href{https://orcid.org/0000-0002-2880-4284}{\includegraphics[scale=0.06]{orcid.pdf}\hspace{1mm}Sanjeev Kalita} \\
	Department of Physics\\
	Gauhati University\\
	Jalukbari, Guwahati-781014, Assam, India \\
	\texttt{sanjeev@gauhati.ac.in} \\
    \And
    \href{https://orcid.org/0000-0003-4301-3496}{\includegraphics[scale=0.06]{orcid.pdf}\hspace{1mm}Debojit Paul} \\
	Department of Physics\\
	Gauhati University\\
	Jalukbari, Guwahati-781014, Assam, India \\
	\texttt{debojit@gauhati.ac.in} \\
    \And
    \href{https://orcid.org/0000-0002-1835-7214}{\includegraphics[scale=0.06]{orcid.pdf}\hspace{1mm}Rajibul Shaikh} \\
	Physics and Applied Mathematics Unit\\
	Indian Statistical Institute\\
	203 Barrackpore Trunk Road, Kolkata-700108,\\
    West Bengal, India \\
	\texttt{rajibulsk@isical.ac.in} \\
}

\date{}

\hypersetup{
pdftitle={A template for the arxiv style},
pdfsubject={q-bio.NC, q-bio.QM},
pdfauthor={David S.~Hippocampus, Elias D.~Striatum},
pdfkeywords={First keyword, Second keyword, More},
}

\begin{document}
\maketitle

\begin{abstract}
	The environment of the Galactic Centre (GC) black hole, Sgr A* gives us new opportunities to test black hole physics and deviation from General Relativity. In this analytical study we calculate null geodesics around the GC black hole for the recently developed rotating black hole metric of f(R) gravity theory. We study the impact of scalaron degree of freedom of f(R) gravity theory on size and shape of the black hole shadow. A lower bound of approximately $10^{-17}$ eV for scalaron mass has been obtained by using $1 \sigma$ upper bounds on deviation parameter measured by the observed shadow size. Scalarons lighter than this bound are found to increase shadow size beyond that measured by the Event Horizon Telescope. We calculate shadow displacement and asymmetry and infer that $10^{-16}$ eV scalarons which produce exact Kerr sized shadow yield 6\% departure from Kerr quadrupole. From asymmetry we infer preservation of black hole no-hair theorem for such scalaron mass and further examine possibility of violation of the theorem. The same mass scale is found to reproduce the PPN parameter ($\gamma$) constrained in the weak field limit of the solar system. Gravitational identifiers, the Kretschmann scalar ($\kappa$) and gravitational potential ($\phi$) have been used to infer scalaron masses in the regime of S-stars near Sgr A* which are found to be consistent with the limits obtained by using shadow scales. We ensure existence of an appropriate general relativistic limit of f(R) gravity scalaron mass in the horizon scale of the black hole.
\end{abstract}


\section{Introduction}

\label{sec:intro}

Several independent cosmological observations indicate that the universe is undergoing accelerated expansion. In the standard model of cosmology, based on General Relativity (GR), a cosmological constant ($\Lambda$) is eligible to drive the late time runaway cosmic expansion. However, this suffers from the Cosmological Constant problem \cite{RevModPhys.61.1} and Cosmic Coincidence problem \cite{2010deto.book.....A}. This motivates search for alternative gravitation theory to account for accelerated expansion of the universe \cite{PhysRevD.70.043528}. Black hole based tests of gravitational theories inform us whether there is a remarkable deviation from GR in the extreme environments of spacetime. Detection of gravitational redshift of star light and Schwarzschild periapsis shift of compact stellar orbits near the Galactic Centre (GC) black hole, Sgr A* \cite{2012Sci...338...84M, 2018A&A...615L..15G}, horizon scale images of Sgr A* and M87* black hole \cite{2019ApJ...875L...1E, 2022ApJ...930L..12E}, constraint on graviton mass \cite{PhysRevLett.125.101102} and test of Hawking's area theorem in gravitational wave observation \cite{2025arXiv250903480T, kw5g-d732} unfold new windows for testing GR and alternative theories of gravity. Remarkable extrapolation of GR  to black hole physics and cosmology is inspired by step by step tests of the theory encompassing various astrophysical scales \cite{Will:2014kxa, 2021MNRAS.502.3761L}.

Cosmologies based on GR involve initial singularity \cite{2008PhR...463..127N, 1980PhLB...91...99S}, mysterious component violating the strong energy condition ($\rho + 3p \geq 0$) required for primordial and late time accelerated expansion \cite{RevModPhys.75.559, 2010deto.book.....A}, the Cosmological Constant problem \cite{RevModPhys.61.1} and illusive, hitherto unidentified Cold Dark Matter particles \cite{1984Natur.311..517B, 1982ApJ...263L...1P, PhysRevD.93.061101}. These are the main reasons behind research on alternatives to GR and their testability. 

Black hole based tests of gravitational physics have important prospects. It is believed that astrophysical black holes are described by the Kerr metric - the stationary, axisymmetric, vacuum solution of Einstein's general relativistic field equations. The higher order multipole moments $(l\geq 2)$ of the external gravitational field of a black hole are determined uniquely by the mass (M) and spin ($a$) of the black hole leading to the no-hair theorem $Q = -Ma^{2}$ where Q is the quadrupole moment(the expression is written in the unit of G = 1 = c). The test of black hole no-hair theorem is essential for understanding asymptotic structure of spacetime and foundation of the theory of gravity. In \cite{PhysRevLett.100.091101} the authors argued that observational evidence of Kerr metric around an astrophysical black hole is not a unique signature of GR. Theories with additional scalar, vector and tensor degree of freedom also possess black holes which are indistinguishable from those of GR. Therefore, a study on deviation from Kerr metric is of potential importance for understanding gravitational physics.

It was expected earlier that orbits of very short period S-stars in the nuclear star cluster near Sgr A* black hole can be used to measure angular momentum ($a$) and quadrupole moment Q of the black hole. But there are potential complications. Detection of such short period stars is associated with perturbation from other compact objects near the black hole \citep{2011CQGra..28v5029S}. The authors in \cite{2010PhRvD..81f2002M} showed that testing no-hair theorem requires stellar orbits as compact as 0.2mpc ($\sim 0.41$ au).

However, discovery of the black hole shadow enclosed by the bright emission rings of gravitationally lensed photons near the M87* black hole and the Sgr A* black hole \cite{2019ApJ...875L...1E, 2022ApJ...930L..12E} has given naive opportunities to test the Kerr hypothesis and the black hole no-hair theorem directly. It is found that size of the black hole shadow is $5M\pm7\%$ \cite{PhysRevLett.125.141104} irrespective of the black hole spin thereby, providing a null test of the Kerr metric hypothesis. On the other hand shape of the black hole shadow is circular except for exceptionally high spin, thanks to the relation $Q = -Ma^2$. Significant asymmetry of the shape for low black hole spin indicates a departure from the no-hair theorem, hence making the shape of the black hole shadow as a probe of no-hair theorem. It is to be noted that the shadow shape is immune to complexities associated with the accretion flow of the black hole and therefore is a clean test of black hole physics \cite{2016ApJ...818..121P}.

New tests provide us with constraints on underlying theories of gravity. These are tests involving large gravitational potential ($\phi \sim M/r$) and large spacetime curvature ($\kappa \sim M/r^3$, the Kretschmann scalar). These tests inform us whether there exists potential deviation from GR in the strong field regimes. It is also possible to map strong field tests of a gravitation theory to its weak field tests. Conventional methods needed to test gravitational theories in the weak field regime of the solar system is to involve theory agnostic Parametrized Post Newtonian (PPN) metric \cite{Will:2014kxa}. PPN parameters are constrained by measuring light deflection near the Sun and in-plane perihelion shift of the planets \cite{PhysRevLett.92.121101, 2003Natur.425..374B, 2014A&A...561A.115V}. These are non linearity of gravity ($\beta$) and spatial curvature produced per unit mass (the Eddington-Robertson-Schiff parameter, $\gamma$). Gravitational tests across various scales provide clue to whether a fundamental theory of gravity is scale invariant. The test of gravity in various scales allows us to leverage the broad conditions that the various scales provide in order to draw conclusions about the theory of gravity that could not have been possible individually \cite{2022ApJ...930L..17E}.

For black hole based tests involving horizon scale imaging observation, one considers theory agnostic non-Kerr metrics free from pathologies \cite{PhysRevD.83.124015, PhysRevD.83.104027, PhysRevD.88.044002, PhysRevD.89.064007, PhysRevD.90.084009, PhysRevD.93.064015}. The parametric deformation of the Kerr metric need not be vacuum solutions of GR. One then constrains the deviations from the Kerr metric. To map strong field tests with the weak field tests, one considers PN expansion of the non-Kerr metric near black holes and takes a conservative hypothesis that weak field constraints on the PPN parameters in the solar system ($\beta -\gamma = 0, \beta = 1 = \gamma$) are valid in the black hole environment \cite{PhysRevLett.125.141104}. Interesting parametrization of non-Kerr metrics are the Johannsen-Psaltis (JP) metric \cite{PhysRevD.83.124015}, Modified Gravity Bumpy Kerr (MGBK) metric \cite{PhysRevD.83.104027} and the Rezolla-Zidenko (RZ) metric \cite{PhysRevD.90.084009}. The EHT imaging of the GC black hole shadow has put correlated constraints on parameters of these metrics which measure departure from the Kerr metric. The constraints on the non-Kerr parameters are found to be weakly dependent on black hole spin.

There are two direct approaches to test an underlying theory of gravity. First, the parametrically deformed Kerr metric is mapped to known black hole solutions of specific alternative gravitational theories such as Randall-Sundram II braneworld  black holes \cite{PhysRevLett.83.4690}, Einstein-Dilaton-Gauss-Bonnet black holes \cite{PhysRevD.83.104002} and dynamical Chern-Simons black holes \cite{PhysRevD.79.084043}. This puts constraint on parameters of those theories. Second, one takes a particular theory of gravity and compares the constraints on parameters of the stationary, axisymmetric and vacuum solution with those measured in various scales. This tells us if the theory has some preferred scale. Here, we adopt the second approach. 

A promising gravitational alternative to GR, which addresses late time cosmic acceleration \cite{2002IJMPD..11..483C, PhysRevD.68.123512, PhysRevD.70.043528, 2010RvMP...82..451S} and flat rotation curve of galaxies \cite{10.1111/j.1365-2966.2007.11401.x} without invoking exotic dark energy and dark matter is the $f(R)$ gravity theory. In \cite{2011PhR...505...59N} the authors studied compatibility of traditional $f(R)$ and Horava-Lifshitz $f(R)$ gravity in realizing unified description of inflation and dark energy. Possibility of realizing bouncing cosmology along with unification of inflation and late time accelerated expansion of the universe in modified gravity theories has also been reviewed \cite{2017PhR...692....1N}.  Here the gravitational field equations are obtained from the Einstein-Hilbert action with Lagrangian of the gravitational field modified as $R \longrightarrow f(R)$, $f$ being a function of the Ricci scalar.
It is written as (in the unit of $G = c=1$)
\begin{equation}
    S = \frac{1}{16\pi}\int\sqrt{-g}\ d^4xf(R) + S_m(\phi_m;g_{\mu\nu}).
\end{equation}
Where $S_m$ is the action for matter fields, $\phi_m$ minimally coupled to spacetime metric, $g_{\mu\nu}$. GR corresponds to $f(R) = R-2\Lambda$, with $\Lambda$ being the cosmological constant.

These theories can originate from time dependent compactification of extra dimensions in string/M+ theory \cite{2003PhLB..576....5N, PhysRevD.74.046004} and curvature corrections to quantum vacuum fluctuations \cite{2018ApJ...855...70K, 2020ApJ...893...31K}. $f(R)$ theories contain an additional scalar field $\psi = f'(R)$ called scalaron and presents a Yukawa correction to the Newtonian potential, with range of the Yukawa force governed by the scalaron mass $M_{\psi} \sim \sqrt{1/\psi'}$ where $\psi' = f''(R)$ (see for example \cite{2018ApJ...855...70K}). 

Earlier observational studies near the GC black hole were useful in constraining Yukawa gravity. \cite{2018ApJ...855...70K, 2020ApJ...893...31K} showed by a theoretical analysis of curvature correction to quantum vacuum fluctuation that Yukawa correction to Newtonian potential naturally arises due to power law ($f(R) = \Sigma_{m>1}^{\infty} a_m R^m$) correction to $\Lambda$CDM like Lagrangian $f(R) = R + \int k^3dk$, where $\int k^3dk$ is the bare cosmological constant given by bare vacuum energy density $\sim k^4$(in the unit of $\hbar = 1=c)$. The Yukawa gravity potential $ ge^{-M_{\psi}r}/r$ ($g$ being the coupling strength) naturally affects astrophysical scales of galaxies and galaxy clusters \cite{2020Univ....6..107D}. The pericentre shift of the S2 star encircling the Sgr A* black hole, detected by GRAVITY Collaboration \cite{2020A&A...636L...5G} has provided us with new opportunity to test gravitational physics near the black hole where gravitational potential is some 2 orders of magnitude larger than that near the Sun.

The authors in \cite{PhysRevLett.118.211101} took help of 19 years of observation of short period S-stars near Sgr A* to constrain strength and scale length of Yukawa fifth force that may arise from unified theory or some models of the dark universe. Adaptive optics (AO) imaging and spectroscopic observation of two stars, S2 and S38 performed by the Keck observatory revealed that there is no deviation from GR for scales within 150 au. They showed that coupling of the fifth force to matter cannot exceed $g \approx 0.01$ at scales comparable to S2-Sgr A* distance. Recent astrometric and spectroscopic data collected through the observation of S2 has been used to constrain interaction strength of the Yukawa fifth force \cite{2025A&A...698L..15G}. The upper bound has been obtained as $|g|<0.003$ for $\lambda \sim 200$ au. It has been reported that for $\lambda < r_{S2}$ ($r_{S2} \sim $ scale of the orbit of S2) no meaningful constraint can be obtained for $g$ by considering equation of motion of the star.

It is certainly indicating a clean region of GR-a wide sub-galactic scale of its validity, without however, a clear reason for why there should be a scale above $\sim 100$ au for modification of gravity. Effect of modified gravity may be hidden in horizon scales which are inaccessible to studies of compact stellar orbits (see for example \cite{2024ApJ...964..127P}). There are other discriminators of gravitational physics. Discovery of bright emission ring near the central brightness depression observed in black hole shadow studies (M87* \cite{2019ApJ...875L...1E} and Sgr A* \cite{2022ApJ...930L..12E}) has shown that distortion of spacetime near the black hole quantified by black hole shadow asymmetry and caused by departure from Kerr metric can be used to test GR based black hole paradigm \cite{PhysRevLett.125.141104, 2010ApJ...718..446J}. Theory agnostic deviation from Kerr metric is parameterized for such studies to examine impact on size and shape of black hole shadows \cite{PhysRevD.83.124015, 2006CQGra..23.4167G}. If the compact object at the GC is a black hole, and there is measurable deviation of the shape of the shadow from circularity, it indicates that perhaps we have a different theory of gravity giving rise to non-Kerr metric in the vacuum limit \cite{2020ApJ...896....7M, PhysRevLett.125.141104}. In this case it would be appropriate if one could recover GR limit of a modified theory of gravity.

Current constraints on Yukawa like fifth force in the GC environment also come from the analysis of shadow cast by the GC black hole \cite{2023CQGra..40p5007V, 2025PDU....4701785N} and measurement of Schwarzschild precession of S2 \cite{2025A&A...698L..15G}. The study of long range fifth force using near earth asteroids has been previously conducted \cite{2023JCAP...04..031T, 2024CmPhy...7..311T}.  \cite{2023CQGra..40p5007V, 2024NatSR..1426932K} constrained interesting extension of GR such as string inspired spacetime, alternative theories of gravity, regular black holes, black hole mimickers, wormhole and naked singularity spacetimes through the EHT shadow of Sgr A*. The study of the M87* black hole with respect to its shape and size has been reported in \cite{2019PhRvD.100d4057B}, where it was found that the Kerr bound on spin is violated.

Earlier studies on compact stellar orbits near Sgr A* \cite{2020ApJ...893...31K, 2023IJMPD..3250021P, 2024ApJ...964..127P} and size of the shadow of Sgr A* \cite{2023EPJC...83..120K, 2024ApJ...964..127P} show that massive scalarons with $M_{\psi} >> M_g$, $M_g$ being the graviton mass ($M_g \sim 10^{-23}$ eV) \cite{PhysRevD.103.122002} reproduce GR like observables. The Lense Thirring effect was explored by \cite{2024ApJ...964..127P}, for the orbit of S2 for scalaron mass range $(10^{-22}-10^{-16})$ eV and it was observed that the scalarons with $10^{-17}$ eV and $10^{-16}$ eV produce Lense Thirring precession of the same size as that of GR for all orbital scales considered. Ultra-light scalarons ($10^{-22}$ eV) however have larger precessions as compared to GR. The GR limit of scalaron mass inspires one to compare it with other scalars of cosmological importance. Scalar particles with mass range $10^{-22}-10^{-10}$ eV constitute ultra-light dark matter component predicted by the axiverse scenario of spectrum of light axion like particles in string theory. \cite{10.1093/mnras/stz2300} predicted capability of the GRAVITY instrument on the VLT to detect scalar dark matter particles of mass range $10^{-20}$-$10^{-18}$ eV near the orbit of S2. It has been shown that scalar particles with $10^{-19}- 10^{-10}$ eV are detectable with gravitational wave detectors \cite{PhysRevD.81.123530}. This is a naive window for scalar field dark matter. We show that scalaron mass compatible with black hole shadow shape and photon rays falls within this window.

Advance of the planetary perihelia and Cassini's bound on the PPN parameter, $\gamma$ in the solar system are found to be satisfied by scalarons which are heavier than graviton by at least 6-7 orders of magnitude \cite{2025PhyS..100f5006P}. 

Several theoretical aspects of black hole physics and  cosmology have been studied by using the shadow as a probe. In \cite{KHAN2020100644} effect of Kerr-Newman black hole surrounded by quintessence dark energy on the shadow shape has been reported. It has been found that frame dragging elongates the shadow in the direction of rotation and charge of the black hole diminishes the shadow. Both spin and charge are found to distort the shadow. The quintessence dark energy has been found to increase the shadow radius. Effect of tidal charge (the electric projection of higher dimensional bulk Weyl tensor on the brane in braneworld gravity theory) on the shadow of Randall -Sundrum braneworld black hole with cosmological constant has been studied in \cite{KHAN2022141}. Constraints on spin and charge of the M87* and the Sgr A* black holes have been obtained in \cite{KHAN2024690} against the EHT data by considering different values of quintessence field around Kerr-Newman black hole. In addition to these properties of event horizon, ergoregion and static limit of braneworld black holes have been studied \cite{KHAN2019100331}. Effect of Scalar-Tensor-Vector (STV) gravity theory on stability of circular orbits of test particles around magnetised Schwarzschild black holes has also been studied \cite{2026EPJC...86..207K}. Rotating Simpson-Visser black hole metric in modified gravity theory has been constructed by \cite{KHAN2026711} with the Newman- Janis algorithm applied to static seed metric of the theory. Calculation of photon geodesics in this spacetime showed that modified gravity coupling parameter increases the black hole shadow radius. The regularization parameter reduces shadow size and enhances asymmetry. Black hole spin has been found to displace shadow in the direction of rotation axis.

In this work we study black hole shadow shape by taking a stationary, axisymmetric, vacuum metric in presence of $f(R)$ gravity scalaron, known as Kerr Scalaron metric developed by \cite{2024ApJ...964..127P}.  We show that $f(R)$ gravity scalaron has an appropriate GR limit while reproducing Kerr like black hole shadow measured by the Event Horizon Telescope (EHT) collaboration. We present scalaron mass which reproduces Kerr like shadow and show that scalarons respect no-hair theorem. We recover the Kerr character of the $f(R)$ gravity metric for an interesting mass scale of the scalaron - that of ultra-light scalar dark matter. We, therefore, reproduce the line of reasoning which ensures that Kerr metric is not unique to GR and exists in $f(R)$ gravity theory \cite{PhysRevLett.100.091101}. It is demonstrated that scalaron mass compatible with GR like shadow matches with the ones derived from different scales such as orbits of the S-stars near Sgr A* \cite{2020ApJ...893...31K, 2023IJMPD..3250021P} and solar system \cite{2025PhyS..100f5006P}. These scalaron masses are found to reproduce solar system like PPN parameter, $\gamma$ in the horizon scale of the black hole. From a novel relation between scalaron mass and black hole mass, we show that scalaron mass compatible with black hole shadow (size and shape) is reproduced by curvature ($\kappa$) and potential ($\phi$) scales experienced by S-stars near the GC black hole. 

The novelty of the finding is in uncovering a general relativistic limit of scalaron mass, a lower mass limit of scalarons on the basis of EHT's available measurement on the Sgr A* black hole shadow, testing black hole no-hair theorem with scalaron mass and linking horizon scale scalaron mass with that obtained earlier in other astrophysical scales.

The paper is organized as follows. In Section \ref{sec2} we calculate the photon orbit in the Kerr Scalaron spacetime and test the no-hair theorem with scalarons by mapping scalaron mass onto quadrupolar correction. In Section \ref{sec3} we reproduce scalaron mass from the gravitational identifiers ($\kappa$, $\phi$). We also justify the scalaron mass on the basis of an interesting observational constraint on the scalaron field amplitude. Section \ref{sec4} presents the results and discussions and section \ref{sec5} concludes. 

\section{Photon orbit with Kerr Scalaron metric and test of no-hair theorem} \label{sec2}
\subsection{The Hamilton-Jacobi formalism}
The stationary, axisymmetric, vacuum solution in presence of f(R) gravity scalaron has been developed by \cite{2024ApJ...964..127P} by applying the Newman-Janis algorithm (NJA) to static, spherically symmetric vacuum solution of f(R) gravity field equations \cite{2018ApJ...855...70K}. It is to be noted that test of spacetime metric through null geodesics does not need field equations \cite{DimitriosPsaltis_2009}. The stationary, axisymmetric, vacuum solution derived by applying NJA to the static, spherically symmetric metric of f (R) gravity theory need not be solution of the specific field equation of the theory (see for example, \cite{PhysRevD.83.124015}). However, as the scalaron mass appears in the new metric (see below), constraint on the mass  automatically puts constraint on the underlying theory. The non-rotating counterpart of the above metric has been derived by \cite{2020ApJ...893...31K} as a static and vacuum solution of the f (R) gravity field equation. The underlying gravity theory is based on gravitational Lagrangian arising from curvature correction to quantum vacuum fluctuations \cite{1968SPhD...12.1040S, 1969JETP...30..372R} near black holes. The Lagrangian is of the form $L \equiv f(R)= \Sigma_{n\geq 0}C_n (k)R^n$, with k being vacuum fluctuation momenta. The Yukawa correction term in the gravitational potential $\frac{-GM}{r} e^{-M_{\psi} (k_{UV},k_{IR} )r}$  is interpreted as quantum gravitational correction to the classical Schwarzschild spacetime, with scalaron mass $M_{\psi}$ being dependent on ultraviolet and infrared cut off scales of vacuum fluctuations. The rotating metric is known as the Kerr Scalaron metric which has the form,
\begin{multline}\label{eq.2}
ds^2 = \left[1-\frac{2Mr}{\rho^2}\left(1+\frac{1}{3}e^{-M_\psi r}\right)\right]dt^2 \\
+ \frac{4Mrasin^2\theta}{\rho^2}\left(1+\frac{1}{3}e^{-M_\psi r}\right)dtd\phi -\frac{\rho^2}{\Delta}dr^2-\rho^2d\theta^2 \\
-\left[r^2+a^2+\frac{2Mra^2sin^2\theta}{\rho^2}\left(1+\frac{1}{3}e^{-M_\psi r}\right)\right]sin^2\theta d\phi^2.
\end{multline}
Here M is the black hole mass and we use the unit system $G = c = 1$, where $G$ is the gravitational constant and $c$ is the speed of light. The quantities $\rho$ and $\Delta$ are defined as
\begin{equation}
   \rho^2 = r^2+a^2\cos^2\theta. 
\end{equation}

\begin{equation} \label{eq4}
   \Delta = r^2-2Mr\left(1+1/3e^{-M_{\psi}r}\right)+a^2. 
\end{equation}

It is to be noted that $M_{\psi}r$ is dimensionless, where $M_{\psi}$ is expressed in eV and $r$ is expressed in au with the conversion $1 au^{-1} = 8.31 \times 10^{-18}$eV. It was shown earlier that power law ($R^m, m>1$) curvature correction to quantum vacuum fluctuation produces a scalaron-black hole mass relationship as $M_\psi \propto \frac{1}{M}$ \cite{2018ApJ...855...70K, 2024JCAP...02..019T}(see section 3). This ensures asymptotic flatness of the Kerr Scalaron metric \cite{2024ApJ...964..127P}. The author in \cite{PhysRevLett.100.091101} demonstrated that if $f(R)$ gravity Lagrangian has the form of a power series 
\begin{equation}
    f(R) = \Sigma_{m=1}^\infty a_mR^m+a_0,
\end{equation}
then vacuum solutions of GR (either R =0 or R = constant) also ensure existence of vacuum solution of $f(R)$ theory of gravity.

The metric presented above is the outcome of field equation residue $\mathcal{R_{\mu\nu}}=G_{\mu\nu}-8\pi T_{\mu\nu}^{(matter)}$ in $f(R)$ gravity theory. The magnitude of field equation residue measures by how much a given metric fails to be the solution of the field equations for a given matter source, $T_{\mu\nu}^{(matter)}$. For vacuum case of Einstein gravity the residue tensor is simply $\mathcal{R_{\mu\nu}}=G_{\mu\nu}=0$ . The $f(R)$ gravity field equations are written as \cite{2018ApJ...855...70K},

\begin{equation}
    \begin{split}
    R_{\mu\nu} -\frac{1}{2}f(R)g_{\mu\nu}+g_{\mu\nu}\nabla^\alpha\nabla_\alpha f'(R)
    - \nabla_\mu\nabla\nu f'(R)=8\pi T^{(matter)}_{\mu\nu}.
    \end{split}
\end{equation}

Therefore, the field equation residue tensor in the vacuum case of $f(R)$ gravity is expressed by the non-zero residue tensor,

\begin{equation}
    \mathcal{R_{\mu\nu}}=G_{\mu\nu}=R_{\mu\nu}-\frac{R}{2}g_{\mu\nu} = 8\pi T^{(eff)}_{\mu\nu}.
\end{equation}

Here $T_{\mu\nu}^{(eff)}$ is the effective energy-momentum tensor \cite{2010RvMP...82..451S} arising from geometrical correction to vacuum field equations. It is obtained as

\begin{equation}
    \begin{split}
    T^{(eff)}_{\mu\nu} = \frac{1}{8\pi}[\nabla_\mu\nabla_\nu f'(R)-g_{\mu\nu}\nabla^\alpha\nabla_\alpha f'(R)
    +\frac{1}{2}g_{\mu\nu}(f(R)-R)].
    \end{split}
\end{equation}

As shown in \cite{2018ApJ...855...70K, 2024JCAP...02..019T} the metric displayed in equation (\ref{eq.2}) is asymptotically flat for $M_\psi \propto 1/M$. Therefore, in this regime the metric approximately becomes the solution of zero residual $\mathcal{R_{\mu\nu}}=0=T_{\mu\nu}^{(eff)}$. Therefore, the metric is a robust approximation for the study of null geodesics and black hole shadow.

In order to see the impact of the Kerr Scalaron metric on shape of the black hole shadow, we proceed by considering the Hamilton-Jacobi (HJ) approach for obtaining the null geodesics around the black hole. The HJ equation is written as
\begin{equation}
    2\frac{\partial S}{\partial \tau} = g^{ij}\frac{\partial S}{\partial x^i}\frac{\partial S}{\partial x^j}.
\end{equation}
Here $\tau$ is the affine parameter and the $g^{ij}$ are the contravariant metric tensor components in the Kerr Scalaron metric. The HJ equation becomes,
\begin{equation}
    \begin{split}
    2\frac{\partial S}{\partial \tau} = \frac{\Sigma^2}{\Delta \rho^2}\left(\frac{\partial S}{\partial t}\right)^2 
    + \frac{4Mar}{\Delta \rho^2}\left(1+\frac{1}{3}e^{-M_{\psi}r}\right)\left(\frac{\partial S}{\partial t}\right)\left(\frac{\partial S}{\partial \phi}\right)-\frac{\Delta-a^2\sin^2\theta}{\rho^2 \Delta \sin^2\theta}\left(\frac{\partial S}{\partial \phi}\right)^2\\\\
    -\frac{\Delta}{\rho^2}\left(\frac{\partial S}{\partial r}\right)^2-\frac{1}{\rho^2}\left(\frac{\partial S}{\partial \theta}\right)^2.
    \end{split}
\end{equation}
On further simplification, we obtain the following convenient form.
\begin{equation}
\begin{split}
    2\frac{\partial S}{\partial \tau} = \frac{1}{\Delta \rho^2}\left[(r^2+a^2)\left(\frac{\partial S}{\partial t}\right)+a\left(\frac{\partial S}{\partial \phi}\right)\right]^2 
    - \frac{1}{\rho^2\sin^2\theta}\left[a \sin^2 \theta \left(\frac{\partial S}{\partial t}\right)+\left(\frac{\partial S}{\partial \phi}\right)\right]^2 \\\\
    -\frac{\Delta}{\rho^2}\left(\frac{\partial S}{\partial r}\right)^2-\frac{1}{\rho^2}\left(\frac{\partial S}{\partial \theta}\right)^2.
\end{split}
\end{equation}

Assuming that the variables are separable, we adopt a principal function of the form 
\begin{equation}
    S = \frac{1}{2}\delta_1 \tau -Et+L_z\phi+S_r(r)+S_\theta(\theta),
\end{equation}

where E and $L_z$ are the particle's energy and angular momentum respectively. $S_r(r)$ and $S_\theta (\theta)$ are the radial and angular principal functions. For the chosen form of $S$, we obtain the separable equations as

\begin{equation}
    \Delta \left(\frac{dS_r}{dr}\right)^2 = \frac{1}{\Delta}\left[(r^2+a^2)E-aL_z\right]^2
    -\left[K + (L_z-aE)^2+\delta_1r^2\right],
\end{equation}
and 
\begin{equation}
    \left(\frac{dS_{\theta}}{d\theta}\right)^2 = K - \left(L_z^2\csc^2\theta-a^2E^2+\delta_1a^2\right)\cos^2\theta.
\end{equation}

Here we abbreviate
\begin{equation}
    R = \left[(r^2+a^2)E-aL_z\right]^2-\Delta \left[K + (L_z-aE)^2+\delta_1r^2\right],
\end{equation}
and 
\begin{equation}
    \Theta = K - \left(L_z^2\csc^2\theta +a^2(\delta_1-E^2)\right)\cos^2\theta.
\end{equation}

\begin{figure*}
\centering
\begin{subfigure}{0.49\textwidth}
    \includegraphics[width=\textwidth]{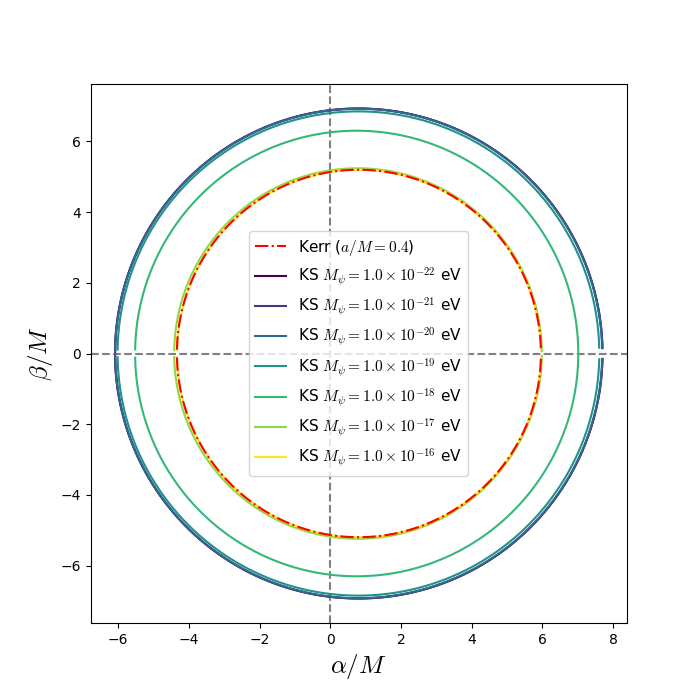}
\end{subfigure}
\hfill
\begin{subfigure}{0.49\textwidth}
    \includegraphics[width=\textwidth]{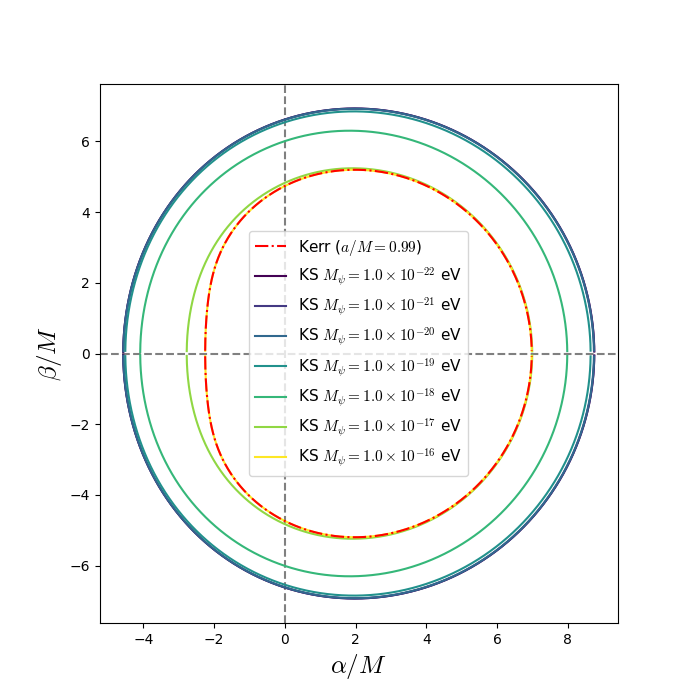}
\end{subfigure}
\caption{The shadow of Kerr Scalaron black hole as observed by a distant observer at $\theta_0 = \pi/2$ for spin \textit{a}/M = 0.4 (left) and \textit{a}/M = 0.99(right). Here KS denotes Kerr Scalaron.}
\label{fig:1}
\end{figure*}

The equations of motion are algebraically simplified to obtain the null geodesics for the variables. For null geodesics, $\delta_1 = 0$ and thus there is a possibility of minimizing the parameters to $\xi = L_z/E$ and $\eta = K/E^2$ which are often called as the impact parameters. Here $K$ denotes the Carter's constant. The equations obtained are as follows:

\begin{equation}
\begin{split}
R = r^4 +r^2(a^2-\xi^2-\eta)
+2Mr[\eta+(\xi-a)^2]\left(1+\frac{1}{3}e^{-M_{\psi}r}\right)-a^2\eta, 
\label{2.15}
\end{split}
\end{equation}
and
\begin{equation}
    \Theta = \eta +a^2\cos^2\theta-\xi^2\cot^2\theta.
\end{equation}

The equation of motion of the radial component of the null geodesics is represented by the function R(r) so that the circular photon orbits satisfies the conditions $R = 0$ and $dR/dr = 0$ (see, for example \cite{1983mtbh.book.....C}). We refrain from displaying the complex algebra involved here. The outcome of the calculations is the following set of impact parameters, 
\begin{equation}
    \xi = \frac{M\left[\alpha'(r^2-a^2)-r(r^2+a^2)\alpha''\right]-r\Delta}{a(r-M\alpha'-mr\alpha'')},
\end{equation}
and 
\begin{equation}
\eta =
\frac{r^3}{
a^2\left[M\alpha'+r(M\alpha''-1)\right]^2}.
\begin{aligned}[t]
\Bigl[&
4Ma^2\alpha'
+6Mr^2\alpha'(1+M\alpha'') \\
&-Mr\left(9M\alpha'^2+4a^2\alpha''\right) 
-r^3(1+M\alpha'')^2
\Bigr].
\end{aligned}
\end{equation}
Here, $\alpha' = 1+\frac{1}{3}e^{-M_\psi r}$, $\alpha'' = \frac{-M_{\psi}e^{-M_{\psi}r}}{3}$ and $\Delta = r^2+a^2-2Mr\alpha'$. These two impact parameters define the celestial coordinates $\alpha$ and $\beta$ of the image as seen by the observer at infinity \cite{1979A&A....75..228L, 2004NCimB.119..489V} as 
\begin{align}
     \alpha = -\xi \csc \theta_0. \\
     \beta = \sqrt{\eta +a^2\csc ^2\theta_0-\xi^2 \cot^2\theta_0}.
\end{align}
Here $\theta_0$ is the inclination angle of the observer. The parametric curve ($(\alpha(\theta), \beta(\theta)$) delineates the shadow boundary as shown in Figure \ref{fig:1}. The shadow morphology for the Kerr Scalaron metric exhibits dependence on the inclination angle of the observer and mass of the scalaron. It is very weakly dependent on spin except for the cases of extremely high spin as shown in the right panel of Figure \ref{fig:1}. For extremely high spin, the presence of scalarons seems to suppress the spin induced distortion, with the shadow approaching the Kerr limit as the scalaron mass approaches $\sim 10^{-16}$ eV. 

Black hole shadow radius is an important observable for testing the nature of gravity. We calculate the average radius of the shadow and its deformation from circularity with the consideration that the shadow obtained is symmetric along the $\alpha$-axis in the $(\alpha, \beta)$ plane. We obtain the geometric centre $(\alpha_c, \beta_c)$ of the shadow by considering $\alpha_c = \int\alpha dA/\int dA$ and $\beta_c = 0$, where $dA$ is the area element. Thus the average radius of the shadow can be obtained by considering a vector from the geometric centre to a point on the boundary and the angle between the $\alpha$ axis and the vector drawn from geometric centre to the periphery \cite{PhysRevD.100.044057, 2023MNRAS.523..375S}. 

\begin{equation}
    R_{sh}^2 = \frac{1}{2\pi}\int_0^{2\pi} l^2(\phi)d\phi.
\end{equation}

Here 
\begin{equation*}
    l(\phi) = \sqrt{(\alpha(\phi)-\alpha_c)^2+\beta(\phi)^2},
\end{equation*}
 and
\begin{equation*}
    \phi = tan^{-1}(\beta(\phi)/(\alpha(\phi)-\alpha_c)).
\end{equation*}

Now we define the deviation from circularity of the shadow as the fractional root mean square distance from the average shadow radius as

\begin{equation}
    \Delta C = \frac{1}{R_{sh}}\sqrt{\frac{1}{2\pi}\int_0^{2\pi}(l(\phi)-R_{sh})^2d\phi}. 
\end{equation}

The shadow radius obtained from the above equation can be  compared with the shadow bound obtained by the EHT observations ($48.7\pm 7 \mu$as) \cite{2022ApJ...930L..12E}. It is now feasible to test impact of scalaron mass on the shadow observables and hence to obtain a general relativistic limit of the f(R) gravity degree of freedom.

\begin{figure*}
\centering
\begin{subfigure}{0.32\textwidth}
    \includegraphics[width=\textwidth]{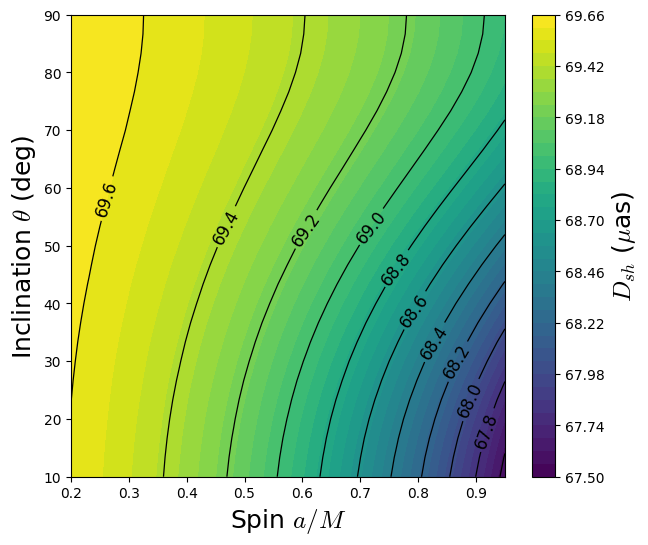}
    \caption{}
    \label{fig:2a}
\end{subfigure}
\hfill
\begin{subfigure}{0.32\textwidth}
    \includegraphics[width=\textwidth]{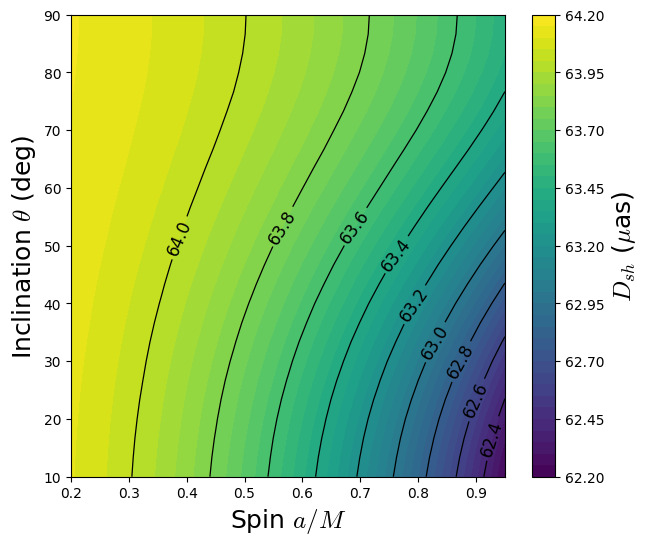}
    \caption{}
    \label{fig:2b}
\end{subfigure}
\hfill
\begin{subfigure}{0.32\textwidth}
    \includegraphics[width=\textwidth]{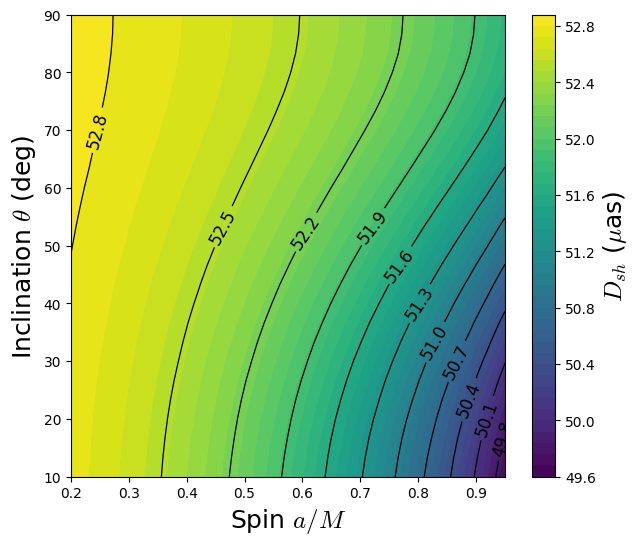}
    \caption{}
    \label{fig:2c}
\end{subfigure}
\caption{Variation of the shadow diameter $D_{sh}$ (in $\mu$as) for inclination angle of the observer $\theta$ and spin $a/M$ for three values of scalaron masses - $10^{-19}$ eV, $10^{-18}$ eV and $10^{-16}$ eV respectively.}
\label{fig:2}
\end{figure*}

\begin{figure*}
\centering
\begin{subfigure}{0.32\textwidth}
    \includegraphics[width=\textwidth]{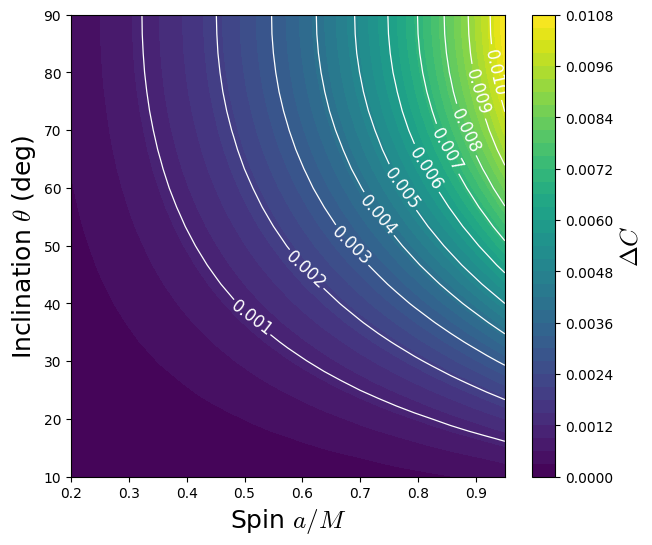}
    \caption{}
    \label{fig:3a}
\end{subfigure}
\hfill
\begin{subfigure}{0.32\textwidth}
    \includegraphics[width=\textwidth]{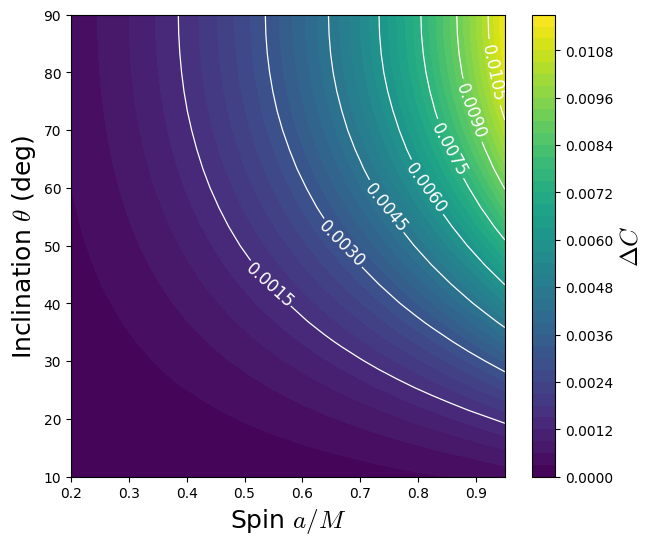}
    \caption{}
    \label{fig:3b}
\end{subfigure}
\hfill
\begin{subfigure}{0.32\textwidth}
    \includegraphics[width=\textwidth]{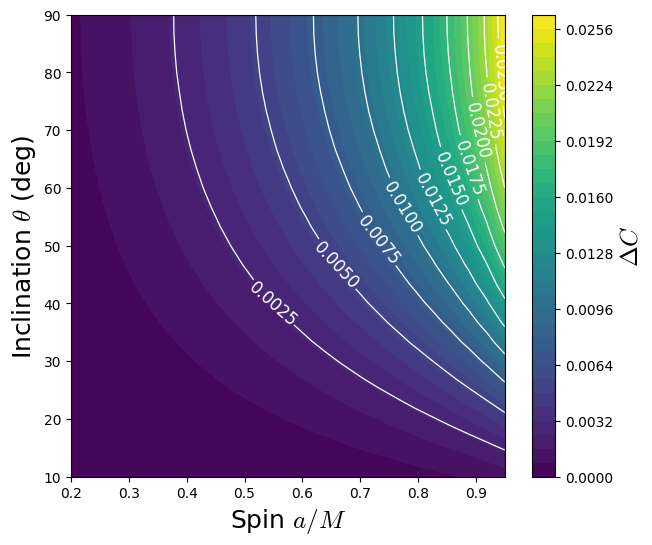}
    \caption{}
    \label{fig:3c}
\end{subfigure}
\caption{Deviation from circularity of the shadow diameter for inclination angle of the observer $\theta$ and spin $a/M$ for three values of scalaron masses - $10^{-19}$ eV, $10^{-18}$ eV and $10^{-16}$eV respectively. }
\label{fig:3}
\end{figure*}

Figure \ref{fig:2} shows shadow diameters plotted with respect to a range of values of black hole spin and observer inclination for three values of scalaron masses, $M_{\psi} = 10^{-19}$ eV, $10^{-18}$ eV and $10^{-16}$ eV. The angular size of the shadow diameter of Sgr A* has been calculated by using the conversion $D_{sh}=2r_{sh}(GM/c^2)$, where $r_{sh}$ is the dimensionless shadow radius calculated in the unit $M=1$ and $GM/c^2 =0.0394$ au, the actual gravitational radius of Sgr A*. The  angular size is then calculated by dividing $D_{sh}$ by distance to Sgr A*. Figure \ref{fig:3} shows deviation from circularity of the shadow for these values of scalaron masses. The maximum deviation from circularity ($\Delta C$) that occurs for $10^{-16}$ eV scalarons is around 0.0256. Therefore, these scalarons produce 2.6\% deviation from circularity. This occurs for very high black hole spin and inclination angle of around $90 \degree$. For lower inclination angle and lower spin deviation reduces further (see Figure \ref{fig:3c}).

\begin{figure*}
\centering
\begin{subfigure}{0.46\textwidth}
    \includegraphics[width=\textwidth]{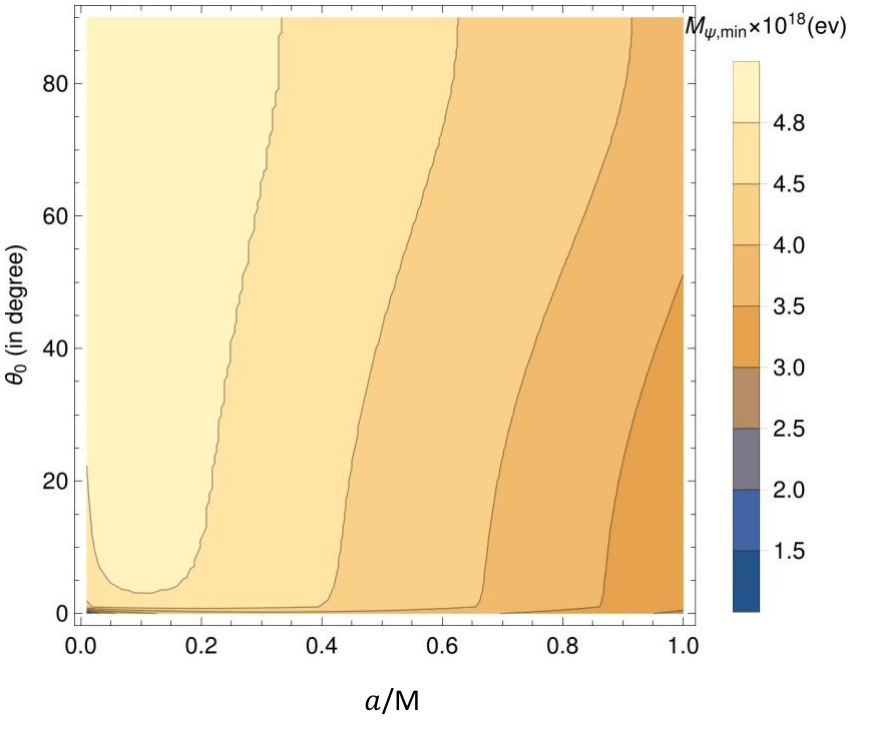}
    \caption{}
    \label{fig:4a}
\end{subfigure}
\hfill
\begin{subfigure}{0.46\textwidth}
    \includegraphics[width=\textwidth]{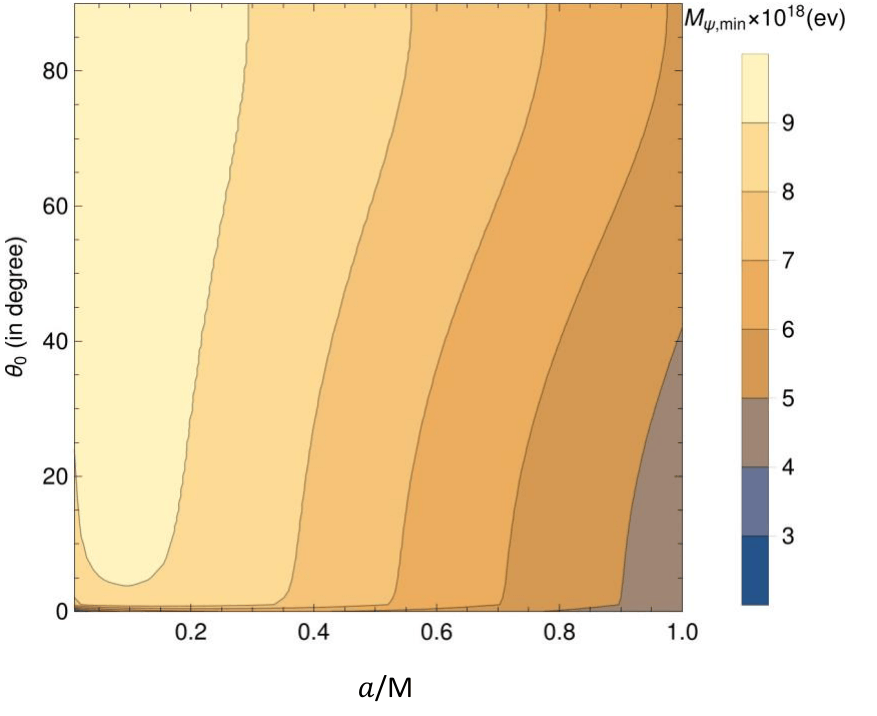}
    \caption{}
    \label{fig:4b}
\end{subfigure}
\caption{Lower bound on scalaron mass as a function of dimensionless spin parameter $a/M$ and angle for upper bounds on deviation parameter reported by Keck (a) and VLT (b) respectively. }
\label{fig:4}
\end{figure*}

We have found (and also apparent from Figure \ref{fig:1}) that, for a given spin and inclination angle, the shadow size increases with decreasing scalaron mass. Below a minimum scalaron mass ($M_{\psi,min} $), the shadow size and hence the deviation parameter $\delta=D_{sh}/D_{sh,Sch}-1$, $D_{sh,Sch}=6\sqrt{3}M$ being the Schwarzschild shadow diameter, go beyond the observed upper limit. Therefore, $M_{\psi,min}$ gives the lower bound on the scalaron mass for a given spin and inclination angle. It is to be noted that, in order to find out the observed value of $\delta$ from the observed angular diameter of the shadow, one needs the mass to distance ratio of Sgr A*. The EHT collaboration takes this value from two other observations, namely the VLT and Keck observation. At the 1 $\sigma$ level, the observed upper bound on the deviation parameter is 0.01 for VLT and 0.05 for Keck \cite{2022ApJ...930L..17E}. The lower bound on scalaron mass has been generated for both the upper bounds.The contour plots for the bound have been shown in Figure \ref{fig:4}.

It has been found that for the VLT upper bound on deviation parameter the minimum scalaron mass lies in the range $2.17\times 10^{-18}$ eV - $9.39\times 10^{-18}$ eV , depending on spin and inclination. On the other hand, the range for the Keck upper bound is  $1.44\times 10^{-18}$ eV  - $4.89\times 10^{-18}$ eV. Irrespective of the spin and inclination angle, the Kerr Scalaron metric satisfies both the observed bounds of the deviation parameter if scalaron mass becomes greater than or equal to $9.39\times 10^{-18}$ eV which is roughly $10^{-17}$ eV.

\subsection{Lensing in Kerr Scalaron metric}

In this section we study the behaviour of null rays in the Kerr and Kerr Scalaron metric. Motion of the null rays and their lensing in Schwarzschild geometry is easily generated \cite{PhysRevD.100.024018}. The practice is to check how the null rays are traced backward to the critical curve of apparent radius (critical impact parameter) $b_c = 3\sqrt{3}M$, below which the rays asymptotically approach the unstable bound orbit at $r = 3M$. 
The method involves solving the geodesic equations for $(r, \phi)$ - the radial and azimuthal coordinates on a Cartesian plane. The effective potential $V_{eff}(r)$ affects the null rays depending on the impact parameter $b$. The effective potential for the Kerr and Kerr Scalaron metric are obtained for the equatorial plane ($\theta = \pi/2$) as

\begin{equation} \label{22}
    V_{eff}^{Kerr} = \frac{1}{r^2}\left[1-\left(\frac{a}{b}\right)^2-\frac{2M}{r}\left(1-\frac{a}{b}\right)^2\right].
\end{equation}

\begin{equation} \label{23}
    V_{eff}^{KS} = \frac{1}{r^2}\left[1-\left(\frac{a}{b}\right)^2-\frac{2M}{r}\left(1+\frac{1}{3}e^{-M_{\psi}r}\right)\left(1-\frac{a}{b}\right)^2\right].
\end{equation}

Here $a$ denotes the spin of the black hole and $b$ is the impact parameter, defined as $b = L/E$.

\begin{figure}
\centering
        \includegraphics[width=0.6\linewidth]{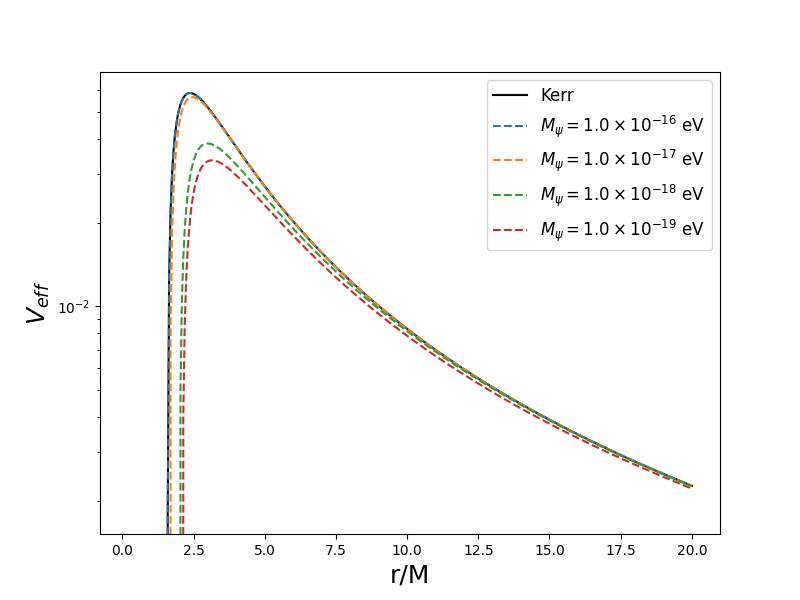}
    \caption{Variation of effective potential for Kerr and Kerr Scalaron metric (see equations \ref{22} and \ref{23}) with respect to $r/M$.}
    \label{fig:5}
\end{figure}

In the rotating case of the behaviour of the null rays is much less tractable as compared to the Schwarzschild counterpart. The trajectories of the prograde and retrograde rays are governed by two impact parameters and the null rays can be described by integrating the geodesic equations. In order to have a clear picture we confine trajectories to the equatorial plane by setting $\theta = \pi/2$. We also consider the Carter's constant to be 0 and proceed by following the evolution of the $(r, \phi)$ geodesics as

\begin{equation} 
    p^r = \pm\frac{\sqrt{r^4 + r^2(a^2 - \xi^2)+2Mr(\xi - a)^2\left(1 + \frac{1}{3}e^{-M_{\psi}r}\right)}}{r^2}, 
\end{equation} 
\begin{equation}
    p^{\phi} = \frac{2Mar\left(1 + \frac{1}{3}e^{-M_{\psi}r}\right) + \left(r^2 - 2Mr\left(1 + \frac{1}{3}e^{-M_{\psi}r}\right)\right) \xi}{r^2\Delta}.
\end{equation}

Here $p^r$ and $p^{\phi}$ are radial and azimuthal momenta of the null rays which depend on all other parameters defined earlier.
    
\begin{figure*}
\centering

\begin{subfigure}{0.46\textwidth}
    \includegraphics[width=\linewidth]{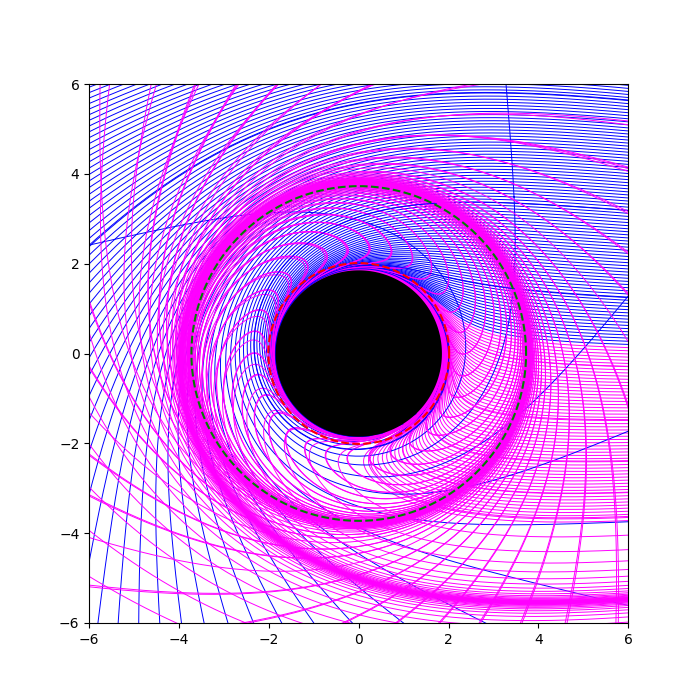}
    \caption{}
\end{subfigure}
\hfill
\begin{subfigure}{0.46\textwidth}
    \includegraphics[width=\linewidth]{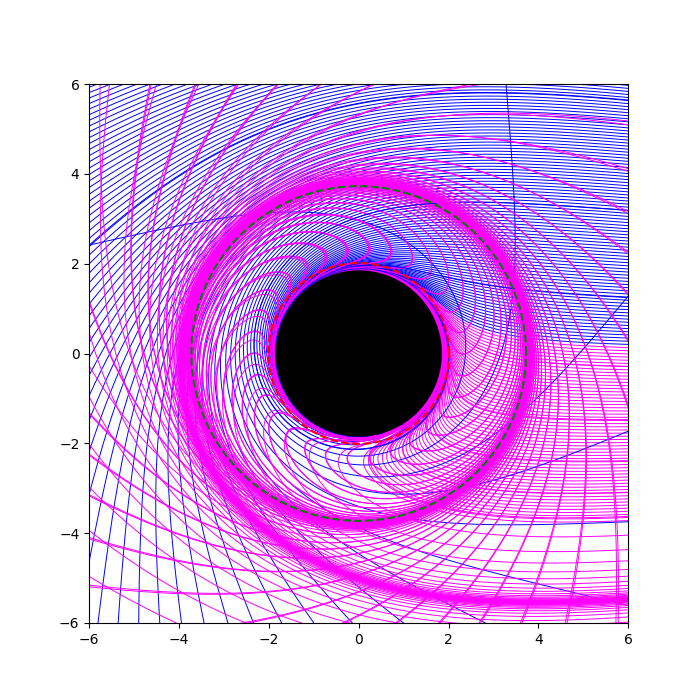}
    \caption{}
\end{subfigure}

\vspace{-1mm}

\begin{subfigure}{0.46\textwidth}
    \includegraphics[width=\linewidth]{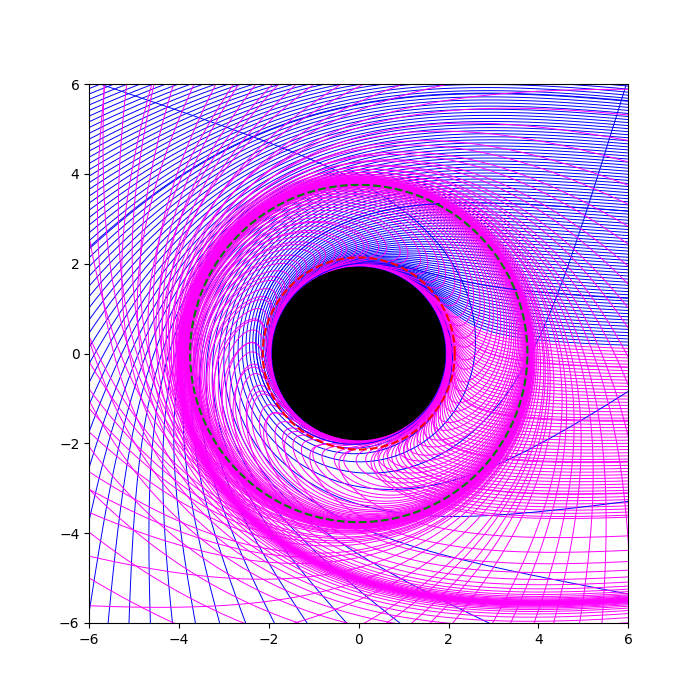}
    \caption{}
\end{subfigure}
\hfill
\begin{subfigure}{0.46\textwidth}
    \includegraphics[width=\linewidth]{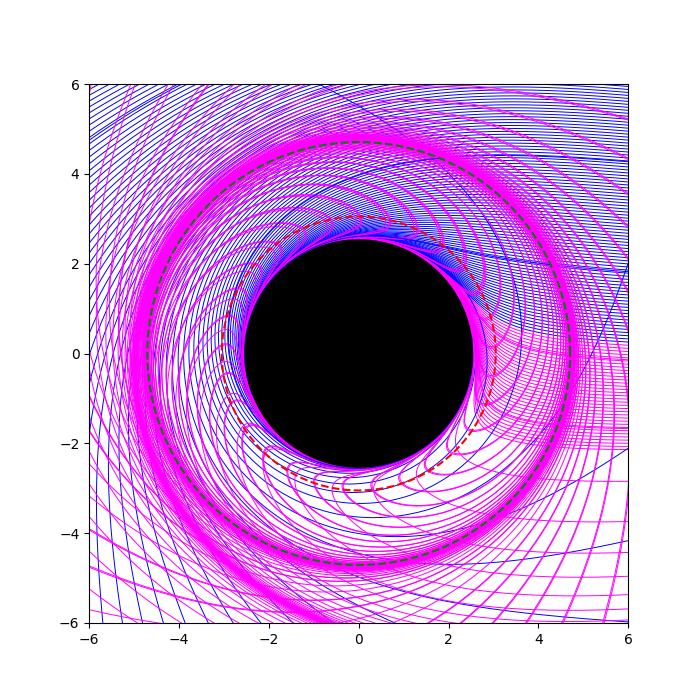}
    \caption{}
\end{subfigure}

\caption{Behaviour of null rays for Kerr spacetime (a) and Kerr Scalaron spacetime ((b), (c) and (d)) with scalaron masses $10^{-16}$ eV, $10^{-17}$ eV, and $10^{-19}$ eV respectively. The prograde and retrograde photon radii are observed whose values have been displayed in Table \ref{table1}}
\label{fig:6}
\end{figure*}

We generate the null ray trajectories for Kerr metric and the Kerr Scalaron metric for black hole spin $a/M =0.7$ and identify scalaron mass which produces a Kerr like pattern of the null rays. Figure \ref{fig:6} shows the change in the null trajectories and we observe that shadow for $10^{-17}$ eV and $10^{-19}$ eV scalarons deviates from the Kerr case.  But for $M_{\psi}=10^{-16}$ eV the shadow geometry becomes indistinguishable from the one predicted by the Kerr metric. This allows us to infer that this scalaron mass represents general relativistic limit of the $f(R)$ gravity theory in the horizon scale.

\begin{table*}
\centering
\resizebox{\textwidth}{!}{
\begin{tabular}{|c|c|c|c|c|}
\hline
Scale of photon trajectories                & Kerr  & Kerr Scalaron ($10^{-16}$ eV) & Kerr Scalaron ($10^{-17}$ eV) & Kerr Scalaron ($10^{-19}$ eV) \\ \hline
Inner horizon/M      & 0.286 & 0.282                         & 0.213                         & 0.199                         \\ \hline
Outer horizon/M     & 1.714 & 1.714                         & 1.802                         & 2.447                         \\ \hline
Prograde radius/M   & 2.013 & 2.013                         & 2.138                         & 3.050                         \\ \hline
Retrograde radius/M  & 3.725 & 3.725                         & 3.755                         & 4.707                         \\ \hline
\end{tabular}
}
\caption{Inner horizon, outer horizon , prograde radius and retrograde radius calculated for Kerr and Kerr Scalaron spacetime.}
\label{table1}
\end{table*}

Table \ref{table1} displays inner horizon, outer horizon, prograde radius and retrograde radius calculated for Kerr and Kerr Scalaron metric. These follow from standard methods. For Kerr spacetime The inner and outer horizon have been calculated by solving the equation $\Delta = r^2- 2Mr+a^2$. The prograde and retrograde radii are obtained by solving the effective potential energy equation for equatorial circular motion of photons and then obtaining the critical radius (see for example \cite{Hobson:2006se}). In the Kerr Scalaron spacetime closed form solutions for these quantities are difficult to obtain due to presence of Yukawa term in the function $\Delta$ (see equation \ref{eq4}). In this case we obtain the values numerically. In both the cases, all the length scales are expressed in dimensionless form by dividing by M. We observe that the shadow size as well as the prograde and retrograde photon radii increase as the scalaron mass decreases.

The author in \cite{2018ApJ...855...70K} reported impact of scalaron mass on light deflection angle at the scale of the very compact stellar orbits near the GC black hole. It was suggested that deviation from GR predicted deflection angle becomes smaller than $0.1 \mu$as at impact parameter of 50 au, if scalarons are massive enough ($M_{\psi} \geq 10^{-18}$ eV). This was a preliminary indication that massive scalaron reproduce GR like results. We reproduce the claim on scalaron mass based on light deflection angle through the scales near the black hole horizon.

\subsection{Test of no-hair theorem} \label{sec:floats}

The black hole no-hair theorem refers to the claim that exterior spacetime of all astrophysical (Kerr) black holes is described only by two independent multipole moments – mass ($M$) and spin ($a$) \cite{PhysRev.164.1776, PhysRevLett.26.331, 1972CMaPh..25..152H}. The quadrupole moment ($Q$) is not an independent degree of freedom. It is expressed in terms of spin as $Q = -Ma^2$. Testing the ”no-hair” theorem requires measuring the three different multipole moments. The size and shape of the black hole shadow are capable of measuring the mass, quadrupole and spin of the black hole and hence test the theorem observationally. We consider the parametrization of the Kerr quadrupole in terms of the parameter $\epsilon$ (called quadrupolar correction) as \cite{2010ApJ...718..446J},

 \begin{equation}
    Q = -M(a^2+\epsilon M^2).
    \label{eq. 12}
\end{equation}

As $\epsilon \longrightarrow 0$ the quadrupole reduces to the Kerr value. The value of this parameter governs strength of departure from the Kerr quadrupole. This relation has been developed for quasi-Kerr metric. By definition, a quasi-Kerr metric refers to the one which is stationary, axisymmetric and vacuum spacetime but differs from the Kerr metric (see \cite{2006CQGra..23.4167G}). For the ease of numerical evaluation it is convenient to define a dimensionless quadrupole moment $q =Q/M^3$ (see \cite{RevModPhys.52.299}) in terms of dimensionless spin parameter $a/M$ as

\begin{equation}
    q = Q/M^3 = -[(a/M)^2 + \epsilon).
\end{equation}

Irrespective of what the theory of gravity is (here, f(R) gravity) one can have parametric extension of the Kerr metric (see e.g. \cite{2020ApJ...896....7M}). That is,
these extensions need not be solutions of Einstein’s field equations. It is true that the quasi-Kerr metric developed in \cite{2006CQGra..23.4167G} is a solution of Einstein’s equation. If correction to Kerr metric is solution of Einstein’s equations, then the central object is not a black hole, rather some exotic star/compact object \cite{2010ApJ...718..446J}. However, if we respect the black hole paradigm or the existence of horizon (as is evident from EHT’s observation) and the spacetime is asymptotically flat (which is the case for KS metric) then it indicates that the horizon scale theory is different from GR. Taking the ansatz- parametric extensions of Kerr metric are not necessarily solutions of GR we adopt the parametrization of quadrupole prescribed in \cite{2010ApJ...718..446J,2006CQGra..23.4167G}.

\begin{figure*}
\centering
\begin{subfigure}{0.46\textwidth}
    \includegraphics[width=\textwidth]{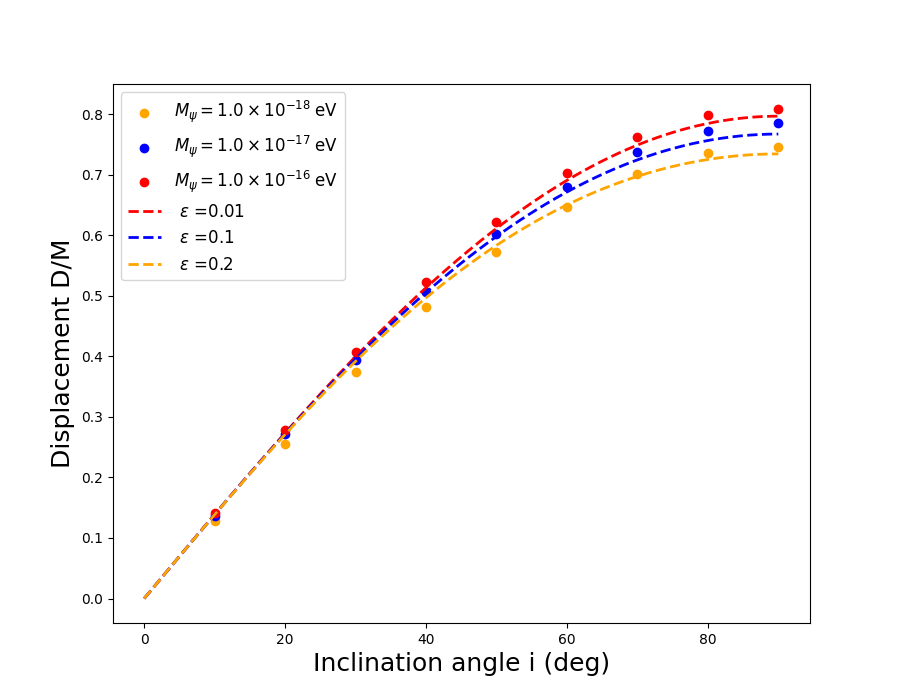}
    \caption{}
    \label{fig:7a}
\end{subfigure}
\hfill
\begin{subfigure}{0.47\textwidth}
    \includegraphics[width=\textwidth]{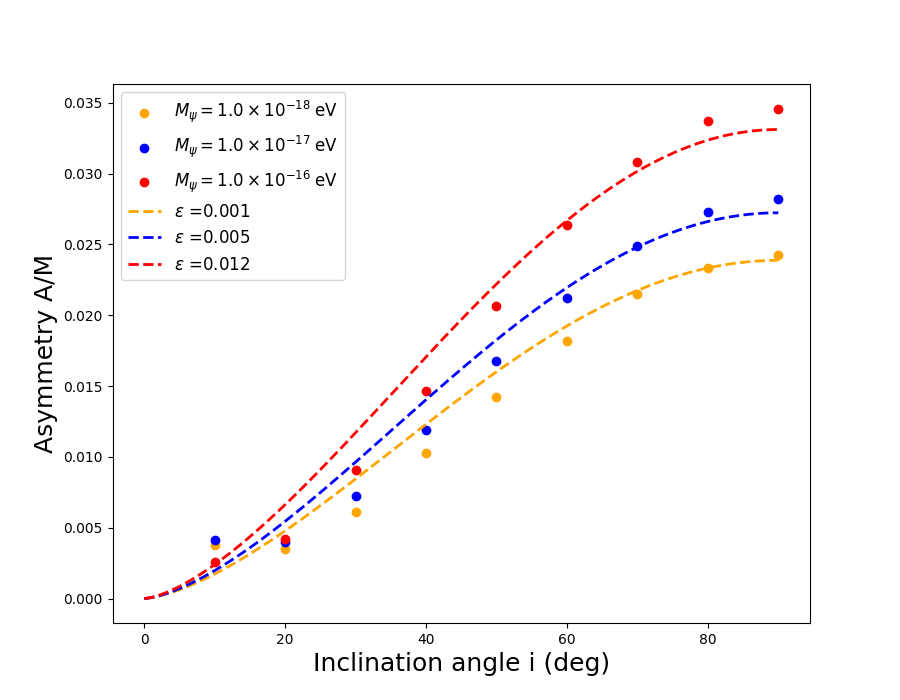}
    \caption{}
    \label{fig:7b}
\end{subfigure}
\caption{Variation of displacement (a) and asymmetry (b) with inclination angle for a fixed value of black hole spin ($a/M = 0.4$) and various scalaron masses.The dashed lines denote the functional form used in \protect\cite{2010ApJ...718..446J} and the $\epsilon$ values are chosen so as to align the scalarons with predicted displacement and asymmetry.}
\label{fig:7}
\end{figure*}

For quantifying dependence of the shape of the shadow on the mass of the scalaron, inclination angle and deviation in the quadrupole moment, we consider the displacement and asymmetry of the black hole shadow. Figure \ref{fig:7}(a) shows the displacement as a function of the inclination angle for a fixed value of black hole spin. We follow the equation \cite{2010ApJ...718..446J}
\begin{equation}\label{eq29}
   \frac{D}{M} = 2\left(\frac{a}{M}\right)\sin i(1-0.41\epsilon \sin^2i).
\end{equation}
Here $D \equiv |x'_{max}+x'_{min}|/2$ is the horizontal displacement of the ring and $i$ is the inclination angle of the disk. The displacement is parameterized by the quadrupolar correction. The displacement is affected by scalaron mass which allows us to recognize a mapping between scalarons and quadrupolar correction. It increases with both increase in inclination angle and mass of the scalaron. A scalaron mass of $10^{-16}$ eV reproduces the displacement for quadrupolar correction $\epsilon = 0.01$. This results in a value, $q = -0.17$. The relative deviation from the Kerr quadrupole predicted by spin $a/M = 0.4$ is, therefore, 6\%.

The degree of asymmetry, quantified by the parameter $A/M$ is considered to be a direct test of the no-hair theorem. Figure \ref{fig:7}(b) illustrates the dependence of the asymmetry parameter on the inclination angle. The following relation for asymmetry is used \cite{2010ApJ...718..446J}
\begin{equation}\label{eq30}
    \frac{A}{M} = \left[0.84\epsilon +0.36\left(\frac{a}{M}\right)^3\right]sin^{3/2}i.
\end{equation}
Like displacement, the asymmetry is also mapped onto the scalaron mass. It may seem that the asymmetry is non negligible and has a high value for higher scalaron mass. However, the magnitude of asymmetry is less as compared to $ 0.36$ which is the maximum asymmetry expected for a Kerr black hole \cite{2010ApJ...718..446J}. It is to be noted here, that while aligning the scalaron mass on displacement and asymmetry, we have respected the condition ($0.0\leq \epsilon \leq 0.5$) required for the parametrization of $D/M$ and $A/M$ (Equation \ref{eq29} and \ref{eq30}). The 6\% departure from the Kerr quadrupole and smaller asymmetries are taken together as benchmark for preservation of black hole ”no-hair” theorem in presence of scalaron. However scalarons lighter than $10^{-16}$ eV such as the minimum bound, $10^{-17}$ eV, may likely produce larger deviation from Kerr quadrupole. This leaves room for violation of the no-hair theorem.

\begin{figure}
\centering
        \includegraphics[width=0.6\linewidth]{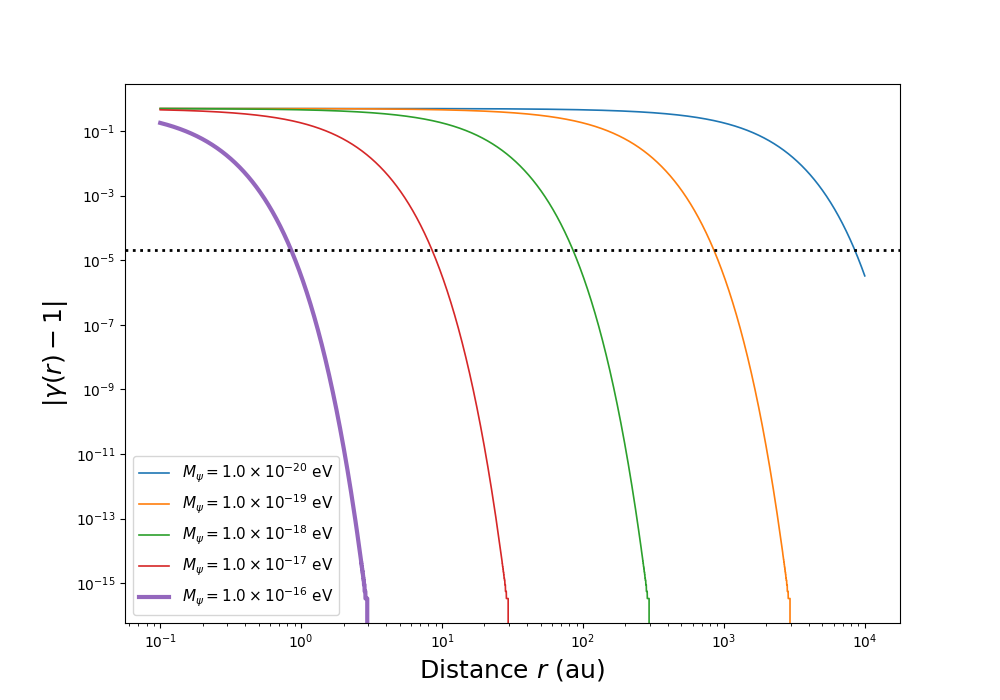}
    \caption{Variation of PPN parameter $\gamma$ with length scale for various scalaron masses in comparison with the solar system bound [$2.1\pm 2.3 \times 10^{-5}$] given by the Cassini mission (the dotted horizontal line) (\protect\cite{2003Natur.425..374B}). The scalaron mass compatible with the general relativistic limit of $f(R)$ theory has been highlighted with the bold purple curve.}
    \label{fig:8}
\end{figure}

Mapping strong field bounds on gravity theory to the weak field bounds is performed by the PPN form of Schwarzschild limit of the Kerr Scalaron metric. The Eddington-Robertson-Schiff parameter $\gamma$ is related to the scalaron mass as \cite{2010deto.book.....A},
\begin{equation}
    \gamma = (3-e^{-M_{\psi}r})/(3+e^{-M_{\psi}r}).
\end{equation}
Whereas low mass scalarons produce substantial departure from GR ($\gamma = 1$), we see that $\gamma -1$ comes down to solar system bound $\gamma-1=(2.1\pm2.3)\times10^{-5}$ given by the Cassini mission \cite{2003Natur.425..374B} within $r = 1 au$ which is fairly close to the scale of shadow diameter of the black hole ($10M \sim 0.4 au$) (see Figure \ref{fig:8}). The values of $\gamma -1$ calculated at different distances from the black hole are displayed in Table \ref{table2}.

\begin{table*}
\centering
\resizebox{\textwidth}{!}{
\begin{tabular}{|c|c|c|c|c|c|}
\hline
r (in au)           & $M_{\psi} = 10^{-20}$ eV & $M_{\psi} = 10^{-19}$ eV & $M_{\psi} = 10^{-18}$ eV & $M_{\psi} = 10^{-17}$ eV & $M_{\psi} = 10^{-16}$ eV \\ \hline
0.1             & $4.9995 \times 10^{-1}$  & $4.9954 \times 10^{-1}$  & $4.9543 \times 10^{-1}$  & $4.5557 \times 10^{-1}$  & $1.7878 \times 10^{-1}$  \\ \hline
1.0             & $4.9954 \times 10^{-1}$  & $4.9543 \times 10^{-1}$  & $4.5557 \times 10^{-1}$  & $1.7878 \times 10^{-1}$  & $3.2708 \times 10^{-6}$  \\ \hline
10              & $4.9543 \times 10^{-1}$  & $4.5557 \times 10^{-1}$  & $1.7878 \time 10^{-1}$   & $3.2708 \times 10^{-6}$  & 0                        \\ \hline
$1 \times 10^2$ & $4.5557 \times 10^{-1}$  & $1.7878 \times 10^{-1}$  & $3.2708 \times 10^{-6}$  & 0                        & 0                        \\ \hline
$1 \times 10^3$ & $1.7878 \times 10^{-1}$  & $3.2708 \times 10^{-6}$  & 0                        & 0                        & 0                        \\ \hline
$1 \times 10^4$ & $3.2708 \times 10^{-6}$  & 0                        & 0                        & 0                        & 0                        \\ \hline
\end{tabular}
}
\caption{Values of $\gamma -1$ calculated at various distances ($r$)  from the black hole (in au) for the scalaron masses $(10^{-16}-10^{-20})$ eV}
\label{table2}
\end{table*}

This confirms that $10^{-16}$ eV scalaron which reproduces Kerr-like observables in black hole shadow is also compatible with the solar system test of GR. From the periapsis shift of S2 star's orbit around the GC black hole it was earlier found that $10^{-16}$ eV scalarons are consistent with GR \cite{2020ApJ...893...31K, 2023IJMPD..3250021P}. Further, using the measurement of EHT deviation parameter, the authors in \cite{2024ApJ...964..127P} demonstrated that scalarons in the mass range $(10^{-17}-10^{-16})$ eV are compatible with the observed size of the emission ring of the GC black hole. Additionally, by studying the advance of perihelia of inner solar system planets, a similar range of scalaron mass was reported by the authors in \cite{2025PhyS..100f5006P}. This exclusively demonstrates the possibility that the strong field bound on gravity theory is compatible with bound arising from weak field. 

\section{Scalaron mass from gravitational identifiers ($\kappa$, $\phi$)}\label{sec3}

\begin{figure}
    \centering
    \includegraphics[width=0.6\linewidth]{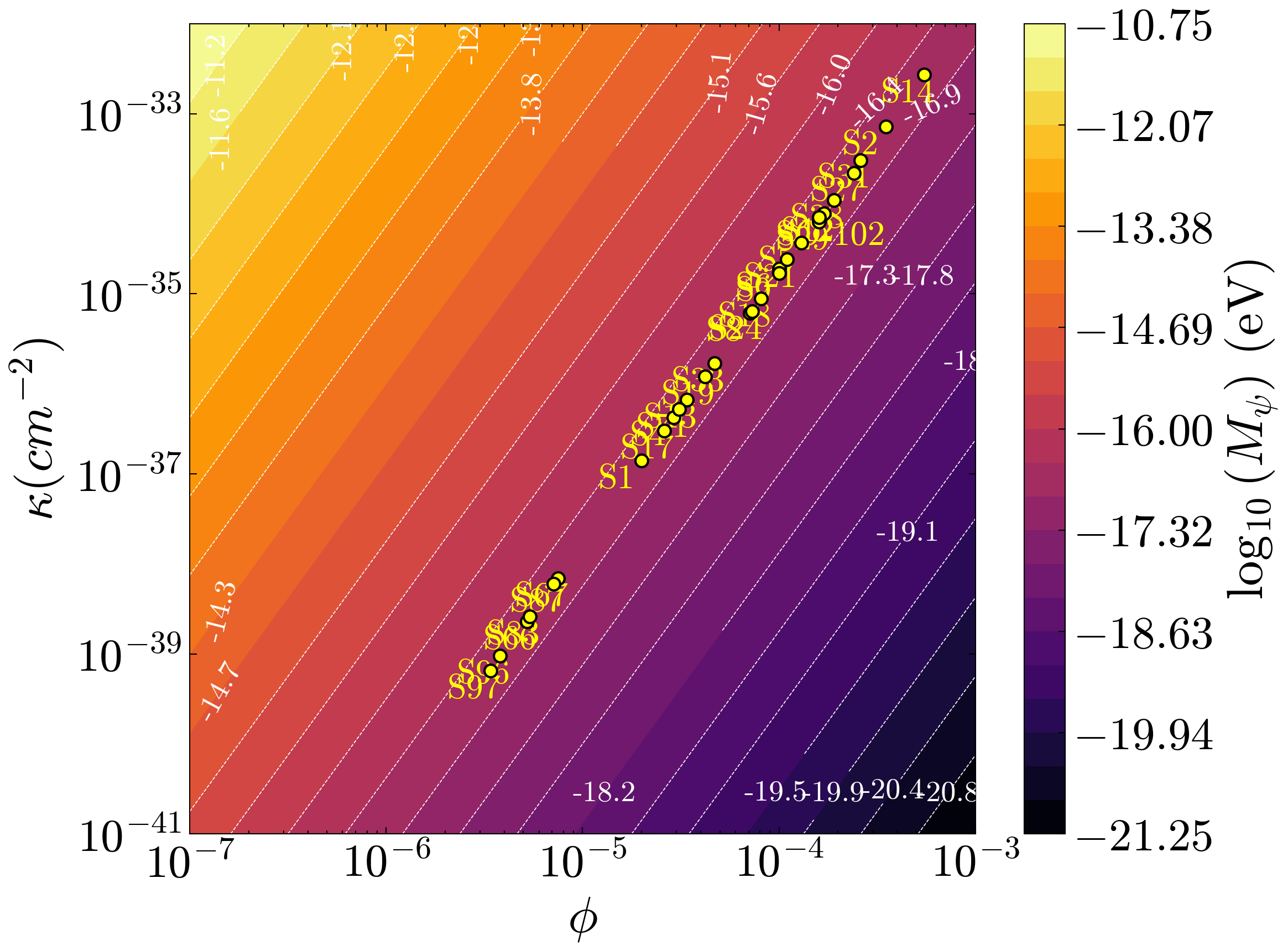}
    \caption{Variation of $M_{\psi}$ for identifiers $\kappa$ and $\phi$. The $\kappa$ and $\phi$ values for S-stars have been adopted from Table 3 of \protect\cite{2015ApJ...802...63B}.}
    \label{fig:9}
\end{figure}

In \cite{2015ApJ...802...63B} the authors identified the Kretschmann curvature ($\kappa$) and gravitational potential ($\phi$) environments where GR has been well tested and is yet to be tested with confidence. They used $\kappa$ and $\phi$ as gravitational identifiers which are related to each other by the relation(in appropriate units of $G$ and $c$) 
\begin{equation}
    \kappa = \sqrt{48} \frac{c^4}{G^2M^2_{BH}}\phi^3.
\end{equation}
Here $M_{BH}$ is the GC black hole mass. The scalaron gravity is a natural consequence of curvature corrected vacuum fluctuations in horizon scale which predicts a novel relation between scalaron mass and black hole mass \cite{2018ApJ...855...70K, 2024JCAP...02..019T}, 
\begin{equation} \label{eqn 33}
    M_{\psi} = 4.30 \times 10^{-10}eV\left(\frac{M_{BH}}{M_\odot}\right)^{-1}.
\end{equation}
This allows us to relate scalaron mass with the gravitational identifiers as

\begin{equation}
    \left(\frac{M_{\psi}}{4.3 \times 10^{-10}eV}\right)^2 = \frac{1}{\sqrt{48}}\left(\frac{\kappa}{\phi^3}\right)\left(\frac{c^4}{G^2M_\odot^2}\right)^{-1}.
\end{equation}
We consider the S-stars near the GC black hole which were considered by \cite{2015ApJ...802...63B} for estimating $\kappa$ and $\phi$ (see Table 3 of \cite{2015ApJ...802...63B}). We have used the range of $\kappa$ and $\phi$ experienced by the S-stars and find a unique value of $M_{\psi} = 10^{-17}$ eV which is the lower bound on scalaron mass permitted by general relativistic observables associated with black hole shadow. This is fairly close to the mass scale $M_{\psi}=10^{-16}$ eV predicted by Equation (\ref{eqn 33}) and the GR limit compatible with black hole shadow size \cite{2024ApJ...964..127P} and shape (the present work). It is displayed in Figure \ref{fig:9}. It has been found that orbits of S-stars and size and shape of the GC black hole are complementary probes of $f(R)$ modified gravity.

We justify the mass scale $M_{\psi} \sim 10^{-17}-10^{-16}$ eV on the basis of an interesting observational bound on the scalaron field amplitude $\delta f'(R_o) = \psi_o-1$ where $\psi_o = f'(R_o)$, $\psi_o = 1$ for GR with $f(R) = R$ and $R_o$ is the background Ricci curvature. The authors in \cite{2013ApJ...779...39J} used distance measurement in the local universe as a new test of gravity theories which obey chameleon screening mechanism \cite{PhysRevLett.93.171104}. They used three distant indicators - cepheids, Tip of the Red Giant Branch (TRGB) and water masers which live in different gravitational environments ($\phi$), with modified gravity being screened upto different extents. The criterion for screening in a spherical mass distribution was given by \cite{PhysRevD.76.064004} as $\phi > \frac{3}{2} \delta f'(R_o)$. It has been found that the background scalaron field values ($\delta f'(R_o)$) above $5\times 10^{-7}$ are ruled out at 95\% confidence level. In this case $\psi_o = 1$ to an incredible precision. Scalaron mass is related to $\psi_o$ and $\psi_o'$ as \cite{2010deto.book.....A},
\begin{equation}\label{eq28}
    M_{\psi}^2 = \frac{Ro}{3}\left(\frac{\psi_o}{\psi_o'R_o}-1\right).
\end{equation}
If deviation from GR is small we can express the theory as
\begin{equation}
    f(R) = R + g(R),
\end{equation}
where $g(R)$ is slowly varying function of R. In exact general relativistic case $g(R) = -2\Lambda$ with $\Lambda$ being the cosmological constant. The background Ricci curvature scales with the $\Lambda$ term as $R_o \sim \Lambda$. Since,

\begin{equation}
    \psi_o' = \left(\frac{d\psi}{dR}\right)_{R_o},
\end{equation}

a Taylor expansion of $\psi(R)$ around $R_o \approx \Lambda \approx 0$ gives

\begin{equation}
    \psi(R)-1= \left(\frac{d\psi}{dR}\right)_{R_o}R.
\end{equation}

This gives,

\begin{equation}
M_{\psi}^2 \approx \frac{1}{3} \left(\frac{R}{\psi-1}\right). 
\end{equation}

Here $\psi-1$ is a very small quantity. Consequently

\begin{equation}
    R \approx 3M_{\psi}^2 \propto 1/M^2_{BH}\propto \frac{\kappa}{\phi^3}.
\end{equation}

This justifies the mass scale $10^{-17}-10^{-16}$ eV.

\section{Results and discussions} \label{sec4}

In this work we have investigated influence of $f(R)$ gravity scalaron degree of freedom on the size and shape of the black hole shadow and hence the black hole no-hair theorem. We generate scalaron mass which produces Kerr like observables in black hole shadow, thereby reporting a lower bound on the scalaron mass. We see the impact of scalarons on null geodesics around the GC black hole and report dependence of shadow size, shape, displacement and asymmetry on scalaron mass. We employ the Kerr Scalaron (KS) metric to sketch the circular photon rings and their shapes. For scalaron mass $M_{\psi} = 10^{-16}$ eV the shadow of the KS black hole coincides with that of the Kerr shadow. On the other hand as the scalarons become lighter, the shadow diameter increases. This occurs due to enhancement of the strength of gravity given by the Yukawa potential ($e^{-M_{\psi}r}/r$). We therefore generate right shadow size compatible with EHT observation with right scalaron mass for the first time. For large spin of the black hole we observe that the distortion is reduced for the low mass scalarons. The lower bound on scalaron mass has been obtained by employing upper bounds on the shadow deviation parameter $\delta$ jointly by the EHT, VLT and Keck observations. It has been found that for the EHT-VLT upper bound on deviation parameter the minimum scalaron mass lies in the range $2.17\times 10^{-18}$ eV - $9.39\times 10^{-18}$ eV , depending on spin and inclination. On the other hand, the range for the EHT-Keck upper bound is  $1.44\times 10^{-18}$ eV  - $4.89\times 10^{-18}$ eV. Irrespective of the spin and inclination angle, the Kerr Scalaron metric satisfies both the observed bounds of the deviation parameter if scalaron mass becomes greater than or equal to $9.39\times 10^{-18}$ eV which is roughly $10^{-17}$ eV. This is therefore realized as the lowest possible scalaron mass giving GR like observables in the shadow scale.

The lensing of null rays around the black hole has also been studied. It has been observed that for scalaron mass of $10^{-16}$ eV, the trajectories reproduces Kerr like shadow. This inference has been found to be spin insensitive (see Figure \ref{fig:1} panels). As the mass of the scalaron decreases, we see a deviation from GR and observe that the size of the black hole shadow increases, the trajectories of the photons differ from those of GR and the unstable photon orbits for both the prograde and retrograde rays, shift. Calculation of the inner and outer horizon and the prograde and retrograde radius in the Kerr Scalaron metric (see Table \ref{table1}) as well as the pattern of null rays in the new spacetime (Figure \ref{fig:6}) clearly show that it is possible to recover a Kerr like metric in f(R) gravity theory for the scalaron mass of $10^{-16}$ eV.

Keeping scalaron mass as a free parameter, we recognize quadrupolar correction which aligns with scalaron masses, reproducing displacement and asymmetry of the rings. The mass scale $10^{-16}$ eV at low black hole spin ($a/M = 0.4$) is found to be compatible with Kerr black hole quadrupole. The scalaron mass of $10^{-16}$ eV reproduces the displacement for quadrupolar correction $\epsilon = 0.01$. This results in a value, $q=-0.17$. The relative deviation from the Kerr quadrupole predicted by black hole spin $a/M=0.4$ is, therefore, 6\%. The maximum asymmetry is still far below the asymmetry allowed by the Kerr hypothesis, $A/M \sim 0.36$. The 6\% departure from the Kerr quadrupole and smaller asymmetries are taken together as benchmark for preservation of black hole no-hair theorem in presence of scalaron. However, scalarons lighter than $10^{-16}$ eV such as the lower bound, $10^{-17}$ eV may likely produce larger deviation from Kerr quadrupole. This leaves room for violation of the no-hair theorem in f(R) gravity theory.

The analysis of the gravitational identifiers has led us to infer that the scalaron mass ($10^{-17}$ eV) scale predicted by these identifiers in the scale of S-stars is compatible with the lower limit of scalaron mass producing Kerr like black hole shadow reported by the EHT observations. We identify $10^{-16}$ eV scalaron mass as an appropriate general relativistic limit of $f(R)$ theory of gravity. Therefore the orbits of S-stars and the size and shape of the GC black hole shadow are complementary probes of $f(R)$ theory of gravity.

The horizon scale of a black hole has given us an appropriate general relativistic limit of scalaron mass. The Yukawa correction ($\frac{g}{3}e^{-M_{\psi}r}/r$) to Newtonian potential at the shadow scale is only 2\% if $g =1$. It further reduces after considering the upper bound on the Yukawa coupling strength, $g$ given by \cite{2025A&A...698L..15G}. It confirms the validity of Birkhoff's theorem in the black hole environment - that the spacetime metric exterior to a spherically symmetric and static body has to be the Schwarzschild (zero spin limit). This also strengthens the claim \cite{PhysRevLett.100.091101} - \textit{the Kerr spacetime is not unique to GR}. The scalaron field is massive. $10^{-16}$ eV corresponds to a Compton wavelength of $0.08$ au, the horizon size of the GC black hole. Gravity gets enhanced below this scale, manifesting as Yukawa fifth force. This clearly ensures validity of GR and GR limit of $f(R)$ gravity theory in astronomical scales such as S2's orbit and black hole shadow.

\section{Conclusion}\label{sec5}

In this study we produce a general relativistic limit of the $f(R)$ gravity theory in the horizon scale of the nearest supermassive black hole. The scalaron mass of $10^{-16}$ eV reproduces general relativistic features of black hole shadow and hence black hole spacetime. Solar system bound on the PPN parameter $\gamma$ is successfully reproduced by $10^{-16}$ eV scalarons. This scalaron mass is tantalizingly close to the one which is compatible with general relativistic periapsis
shift of S2 star \cite{2020ApJ...893...31K} , size of the GC black hole shadow (this work) and advance of perihelia of planets in the solar system \cite{2025PhyS..100f5006P}. The scalaron mass scale reported here is also compatible with the mass scale of the Yukawa fifth force mediator and ultra-light dark matter ($10^{-21}$ - $10^{-15}$ eV) coming from asteroid astrometric data and $10^{-18}$ - $10^{-17}$ eV coming from analysis of the orbit of the asteroid Bennu \cite{2023JCAP...04..031T, 2024CmPhy...7..311T}. Similarity of the horizon scale scalaron mass with that realized in the solar system is a non-trivial outcome. We conclude with an exciting possibility that general relativistic limit of $f(R)$ gravity theory is likely to be scale invariant. The scalaron mass scale presented here is likely to give us a new understanding of gravitational physics in a new regime of spacetime.

Future gravitational wave observations can be used to test scalaron gravity through supermassive black hole merger. In the vicinity of compact objects, the coupling between gravitational waves and matter and modified dispersion relation of the gravitational waves due to massive gravity modes are two probes for strong field nature of gravity. The deviation of modified theories of gravity from GR becomes appreciable at low frequencies due to the presence of the massive modes of the theory. The deviation of results as compared to GR are non negligible for compact objects acting as lenses for gravitational waves. For example, LISA's operating frequency band $\sim$ $10^{-4}$ Hz - $1$ Hz is eligible to detect $10^{-19}-10^{-15}$ eV scalarons through lensing of gravitational wave by $10^3-10^6 M_\odot$ compact objects \cite{SHARMA2024139093}. 

We wish to draw attention to few points of unusual interest which carry potential for future studies on gravity theories beyond GR through astronomical observables.

\begin{itemize}
    \item Existence of a GR limit of $f(R)$ scalaron gravity indicates that the theory can act as a close alternative to GR in cosmological setting. 
    \item Scalarons of the mass scale presented here can interestingly affect gravitational wave propagation and can be tested via future low frequency gravitational wave observatories.
    \item The GR limit of scalaron mass phenomenologically strengthens the claim - \textit{the Kerr metric is not unique to GR}. 
    \item The GR limit of scalaron mass coincides with ultra-light scalar field dark matter. This may 
    be of use for investigating a new class of dark matter particles in the context of an interesting alternative to GR.
    \item The Compton wavelength of the $10^{-16}$ eV scalarons ($\sim 0.08$ au) matches with the size of extra dimensions (5 dimensional anti de-Sitter space) constrained by EHT measurements \cite{PhysRevD.110.064079}.The bound is $10^{-2}$ au. This is an important outcome for future astronomical observations to be employed for testing new gravitational physics and effect of extra dimensions in an environment accessible to us.
    
\end{itemize}

\bibliographystyle{unsrtnat}
\bibliography{references}  

\begin{thebibliography}{93}
\providecommand{\natexlab}[1]{#1}
\providecommand{\url}[1]{\texttt{#1}}
\expandafter\ifx\csname urlstyle\endcsname\relax
  \providecommand{\doi}[1]{doi: #1}\else
  \providecommand{\doi}{doi: \begingroup \urlstyle{rm}\Url}\fi

\bibitem[Weinberg(1989)]{RevModPhys.61.1}
Steven Weinberg.
\newblock The cosmological constant problem.
\newblock \emph{Rev. Mod. Phys.}, 61:\penalty0 1--23, Jan 1989.
\newblock \doi{10.1103/RevModPhys.61.1}.
\newblock URL \url{https://link.aps.org/doi/10.1103/RevModPhys.61.1}.

\bibitem[{Amendola} and {Tsujikawa}(2010)]{2010deto.book.....A}
Luca {Amendola} and Shinji {Tsujikawa}.
\newblock \emph{{Dark Energy: Theory and Observations}}.
\newblock Cambridge University Press, Cambridge, UK, 2010.
\newblock URL \url{https://ui.adsabs.harvard.edu/abs/2010deto.book.....A}.

\bibitem[Carroll et~al.(2004)Carroll, Duvvuri, Trodden, and Turner]{PhysRevD.70.043528}
Sean~M. Carroll, Vikram Duvvuri, Mark Trodden, and Michael~S. Turner.
\newblock Is cosmic speed-up due to new gravitational physics?
\newblock \emph{Phys. Rev. D}, 70:\penalty0 043528, Aug 2004.
\newblock \doi{10.1103/PhysRevD.70.043528}.
\newblock URL \url{https://link.aps.org/doi/10.1103/PhysRevD.70.043528}.

\bibitem[{Meyer} et~al.(2012){Meyer}, {Ghez}, {Sch{\"o}del}, {Yelda}, {Boehle}, {Lu}, {Do}, {Morris}, {Becklin}, and {Matthews}]{2012Sci...338...84M}
L.~{Meyer}, A.~M. {Ghez}, R.~{Sch{\"o}del}, S.~{Yelda}, A.~{Boehle}, J.~R. {Lu}, T.~{Do}, M.~R. {Morris}, E.~E. {Becklin}, and K.~{Matthews}.
\newblock The shortest-known-period star orbiting our galaxy{\textquoteright}s supermassive black hole.
\newblock \emph{Science}, 338\penalty0 (6103):\penalty0 84, October 2012.
\newblock \doi{10.1126/science.1225506}.
\newblock URL \url{https://ui.adsabs.harvard.edu/abs/2012Sci...338...84M}.

\bibitem[{Abuter} et~al.(2018)]{2018A&A...615L..15G}
R.~{Abuter} et~al.
\newblock Detection of the gravitational redshift in the orbit of the star s2 near the galactic centre massive black hole.
\newblock \emph{Astron. Astrophys.}, 615:\penalty0 L15, July 2018.
\newblock \doi{10.1051/0004-6361/201833718}.
\newblock URL \url{https://ui.adsabs.harvard.edu/abs/2018A&A...615L..15G}.

\bibitem[{Akiyama} et~al.(2019)]{2019ApJ...875L...1E}
Kazunori {Akiyama} et~al.
\newblock First {M87} event horizon telescope results. i. the shadow of the supermassive black hole.
\newblock \emph{Astrophys. J. Lett}, 875\penalty0 (1):\penalty0 L1, April 2019.
\newblock \doi{10.3847/2041-8213/ab0ec7}.
\newblock URL \url{https://ui.adsabs.harvard.edu/abs/2019ApJ...875L...1E}.

\bibitem[{Akiyama} et~al.(2022{\natexlab{a}})]{2022ApJ...930L..12E}
Kazunori {Akiyama} et~al.
\newblock First {Sagittarius A*} event horizon telescope results. i. the shadow of the supermassive black hole in the center of the milky way.
\newblock \emph{Astrophys. J. Lett}, 930\penalty0 (2):\penalty0 L12, May 2022{\natexlab{a}}.
\newblock \doi{10.3847/2041-8213/ac6674}.
\newblock URL \url{https://ui.adsabs.harvard.edu/abs/2022ApJ...930L..12E}.

\bibitem[Abbott et~al.(2020)]{PhysRevLett.125.101102}
R.~Abbott et~al.
\newblock {GW190521}: A binary black hole merger with a total mass of $150\text{ }\text{ }{M}_{\ensuremath{\bigodot}}$.
\newblock \emph{Phys. Rev. Lett}, 125:\penalty0 101102, Sep 2020.
\newblock \doi{10.1103/PhysRevLett.125.101102}.
\newblock URL \url{https://link.aps.org/doi/10.1103/PhysRevLett.125.101102}.

\bibitem[{Tang} et~al.(2025){Tang}, {Wang}, {Li}, and {Fan}]{2025arXiv250903480T}
Shao-Peng {Tang}, Hai-Tian {Wang}, Yin-Jie {Li}, and Yi-Zhong {Fan}.
\newblock {Verification of the Black Hole Area Law with GW230814}.
\newblock \emph{arXiv e-prints}, art. arXiv:2509.03480, September 2025.
\newblock \doi{10.48550/arXiv.2509.03480}.
\newblock URL \url{https://ui.adsabs.harvard.edu/abs/2025arXiv250903480T}.

\bibitem[Abac et~al.(2025)]{kw5g-d732}
A.~G. Abac et~al.
\newblock {GW250114}: Testing hawking's area law and the kerr nature of black holes.
\newblock \emph{Phys. Rev. Lett}, 135:\penalty0 111403, Sep 2025.
\newblock \doi{10.1103/kw5g-d732}.
\newblock URL \url{https://link.aps.org/doi/10.1103/kw5g-d732}.

\bibitem[Will(2014)]{Will:2014kxa}
Clifford~M. Will.
\newblock The confrontation between general relativity and experiment.
\newblock \emph{Living Rev. Relativ.}, 17:\penalty0 4, 2014.
\newblock \doi{10.12942/lrr-2014-4}.

\bibitem[{Lalremruati} and {Kalita}(2021)]{2021MNRAS.502.3761L}
P.~C. {Lalremruati} and Sanjeev {Kalita}.
\newblock Periastron shift of compact stellar orbits from general relativistic and tidal distortion effects near {Sgr A*}.
\newblock \emph{Mon. Not. R. Astron. Soc.}, 502\penalty0 (3):\penalty0 3761--3768, April 2021.
\newblock \doi{10.1093/mnras/stab129}.
\newblock URL \url{https://ui.adsabs.harvard.edu/abs/2021MNRAS.502.3761L}.

\bibitem[{Novello} and {Bergliaffa}(2008)]{2008PhR...463..127N}
M.~{Novello} and S.~E.~Perez {Bergliaffa}.
\newblock Bouncing cosmologies.
\newblock \emph{Phys. Rep}, 463\penalty0 (4):\penalty0 127--213, July 2008.
\newblock \doi{10.1016/j.physrep.2008.04.006}.
\newblock URL \url{https://ui.adsabs.harvard.edu/abs/2008PhR...463..127N}.

\bibitem[{Starobinsky}(1980)]{1980PhLB...91...99S}
A.~A. {Starobinsky}.
\newblock A new type of isotropic cosmological models without singularity.
\newblock \emph{Phys. Lett. B}, 91\penalty0 (1):\penalty0 99--102, March 1980.
\newblock \doi{10.1016/0370-2693(80)90670-X}.
\newblock URL \url{https://ui.adsabs.harvard.edu/abs/1980PhLB...91...99S}.

\bibitem[Peebles and Ratra(2003)]{RevModPhys.75.559}
P.~J.~E. Peebles and Bharat Ratra.
\newblock The cosmological constant and dark energy.
\newblock \emph{Rev. Mod. Phys.}, 75:\penalty0 559--606, Apr 2003.
\newblock \doi{10.1103/RevModPhys.75.559}.
\newblock URL \url{https://link.aps.org/doi/10.1103/RevModPhys.75.559}.

\bibitem[{Blumenthal} et~al.(1984){Blumenthal}, {Faber}, {Primack}, and {Rees}]{1984Natur.311..517B}
G.~R. {Blumenthal}, S.~M. {Faber}, J.~R. {Primack}, and M.~J. {Rees}.
\newblock Formation of galaxies and large-scale structure with cold dark matter.
\newblock \emph{Nature}, 311:\penalty0 517--525, October 1984.
\newblock \doi{10.1038/311517a0}.
\newblock URL \url{https://ui.adsabs.harvard.edu/abs/1984Natur.311..517B}.

\bibitem[{Peebles}(1982)]{1982ApJ...263L...1P}
P.~J.~E. {Peebles}.
\newblock Large-scale background temperature and mass fluctuations due to scale-invariant primeval perturbations.
\newblock \emph{Astrophys. J. Lett.}, 263:\penalty0 L1--L5, December 1982.
\newblock \doi{10.1086/183911}.
\newblock URL \url{https://ui.adsabs.harvard.edu/abs/1982ApJ...263L...1P}.

\bibitem[Amole et~al.(2016)]{PhysRevD.93.061101}
C.~Amole et~al.
\newblock Improved dark matter search results from {PICO-2L} run 2.
\newblock \emph{Phys. Rev. D}, 93:\penalty0 061101, Mar 2016.
\newblock \doi{10.1103/PhysRevD.93.061101}.
\newblock URL \url{https://link.aps.org/doi/10.1103/PhysRevD.93.061101}.

\bibitem[Psaltis et~al.(2008)Psaltis, Perrodin, Dienes, and Mocioiu]{PhysRevLett.100.091101}
Dimitrios Psaltis, Delphine Perrodin, Keith~R. Dienes, and Irina Mocioiu.
\newblock Kerr black holes are not unique to general relativity.
\newblock \emph{Phys. Rev. Lett.}, 100:\penalty0 091101, Mar 2008.
\newblock \doi{10.1103/PhysRevLett.100.091101}.
\newblock URL \url{https://link.aps.org/doi/10.1103/PhysRevLett.100.091101}.

\bibitem[{Sadeghian} and {Will}(2011)]{2011CQGra..28v5029S}
Laleh {Sadeghian} and Clifford~M. {Will}.
\newblock {Testing the black hole no-hair theorem at the galactic center: perturbing effects of stars in the surrounding cluster}.
\newblock \emph{Class. Quantum Gravity}, 28\penalty0 (22):\penalty0 225029, November 2011.
\newblock \doi{10.1088/0264-9381/28/22/225029}.
\newblock URL \url{https://ui.adsabs.harvard.edu/abs/2011CQGra..28v5029S}.

\bibitem[{Merritt} et~al.(2010){Merritt}, {Alexander}, {Mikkola}, and {Will}]{2010PhRvD..81f2002M}
David {Merritt}, Tal {Alexander}, Seppo {Mikkola}, and Clifford~M. {Will}.
\newblock {Testing properties of the Galactic center black hole using stellar orbits}.
\newblock \emph{Phys. Rev. D}, 81\penalty0 (6):\penalty0 062002, March 2010.
\newblock \doi{10.1103/PhysRevD.81.062002}.
\newblock URL \url{https://ui.adsabs.harvard.edu/abs/2010PhRvD..81f2002M}.

\bibitem[Psaltis et~al.(2020)]{PhysRevLett.125.141104}
Dimitrios Psaltis et~al.
\newblock Gravitational test beyond the first post-newtonian order with the shadow of the {M87} black hole.
\newblock \emph{Phys. Rev. Lett}, 125:\penalty0 141104, Oct 2020.
\newblock \doi{10.1103/PhysRevLett.125.141104}.
\newblock URL \url{https://link.aps.org/doi/10.1103/PhysRevLett.125.141104}.

\bibitem[{Psaltis} et~al.(2016){Psaltis}, {Wex}, and {Kramer}]{2016ApJ...818..121P}
Dimitrios {Psaltis}, Norbert {Wex}, and Michael {Kramer}.
\newblock \emph{Astrophys. J.}, 818\penalty0 (2):\penalty0 121, February 2016.
\newblock \doi{10.3847/0004-637X/818/2/121}.
\newblock URL \url{https://ui.adsabs.harvard.edu/abs/2016ApJ...818..121P}.

\bibitem[Shapiro et~al.(2004)Shapiro, Davis, Lebach, and Gregory]{PhysRevLett.92.121101}
S.~S. Shapiro, J.~L. Davis, D.~E. Lebach, and J.~S. Gregory.
\newblock Measurement of the solar gravitational deflection of radio waves using geodetic very-long-baseline interferometry data, 1979--1999.
\newblock \emph{Phys. Rev. Lett}, 92:\penalty0 121101, Mar 2004.
\newblock \doi{10.1103/PhysRevLett.92.121101}.
\newblock URL \url{https://link.aps.org/doi/10.1103/PhysRevLett.92.121101}.

\bibitem[{Bertotti} et~al.(2003){Bertotti}, {Iess}, and {Tortora}]{2003Natur.425..374B}
B.~{Bertotti}, L.~{Iess}, and P.~{Tortora}.
\newblock A test of general relativity using radio links with the cassini spacecraft.
\newblock \emph{Nature}, 425\penalty0 (6956):\penalty0 374--376, September 2003.
\newblock \doi{10.1038/nature01997}.
\newblock URL \url{https://ui.adsabs.harvard.edu/abs/2003Natur.425..374B}.

\bibitem[{Verma} et~al.(2014){Verma}, {Fienga}, {Laskar}, {Manche}, and {Gastineau}]{2014A&A...561A.115V}
A.~K. {Verma}, A.~{Fienga}, J.~{Laskar}, H.~{Manche}, and M.~{Gastineau}.
\newblock Use of messenger radioscience data to improve planetary ephemeris and to test general relativity.
\newblock \emph{Astron. Astrophys.}, 561:\penalty0 A115, January 2014.
\newblock \doi{10.1051/0004-6361/201322124}.
\newblock URL \url{https://ui.adsabs.harvard.edu/abs/2014A&A...561A.115V}.

\bibitem[{Akiyama} et~al.(2022{\natexlab{b}})]{2022ApJ...930L..17E}
Kazunori {Akiyama} et~al.
\newblock First {Sagittarius A* Event Horizon Telescope Results. VI.} {Testing the Black Hole M}etric.
\newblock \emph{Astrophys. J. Lett.}, 930\penalty0 (2):\penalty0 L17, May 2022{\natexlab{b}}.
\newblock \doi{10.3847/2041-8213/ac6756}.
\newblock URL \url{https://ui.adsabs.harvard.edu/abs/2022ApJ...930L..17E}.

\bibitem[Johannsen and Psaltis(2011)]{PhysRevD.83.124015}
Tim Johannsen and Dimitrios Psaltis.
\newblock Metric for rapidly spinning black holes suitable for strong-field tests of the no-hair theorem.
\newblock \emph{Phys. Rev. D}, 83:\penalty0 124015, Jun 2011.
\newblock \doi{10.1103/PhysRevD.83.124015}.
\newblock URL \url{https://link.aps.org/doi/10.1103/PhysRevD.83.124015}.

\bibitem[Vigeland et~al.(2011)Vigeland, Yunes, and Stein]{PhysRevD.83.104027}
Sarah Vigeland, Nicol\'as Yunes, and Leo~C. Stein.
\newblock Bumpy black holes in alternative theories of gravity.
\newblock \emph{Phys. Rev. D}, 83:\penalty0 104027, May 2011.
\newblock \doi{10.1103/PhysRevD.83.104027}.
\newblock URL \url{https://link.aps.org/doi/10.1103/PhysRevD.83.104027}.

\bibitem[Johannsen(2013)]{PhysRevD.88.044002}
Tim Johannsen.
\newblock Regular black hole metric with three constants of motion.
\newblock \emph{Phys. Rev. D}, 88:\penalty0 044002, Aug 2013.
\newblock \doi{10.1103/PhysRevD.88.044002}.
\newblock URL \url{https://link.aps.org/doi/10.1103/PhysRevD.88.044002}.

\bibitem[Cardoso et~al.(2014)Cardoso, Pani, and Rico]{PhysRevD.89.064007}
Vitor Cardoso, Paolo Pani, and Jo\~ao Rico.
\newblock On generic parametrizations of spinning black-hole geometries.
\newblock \emph{Phys. Rev. D}, 89:\penalty0 064007, Mar 2014.
\newblock \doi{10.1103/PhysRevD.89.064007}.
\newblock URL \url{https://link.aps.org/doi/10.1103/PhysRevD.89.064007}.

\bibitem[Rezzolla and Zhidenko(2014)]{PhysRevD.90.084009}
Luciano Rezzolla and Alexander Zhidenko.
\newblock New parametrization for spherically symmetric black holes in metric theories of gravity.
\newblock \emph{Phys. Rev. D}, 90:\penalty0 084009, Oct 2014.
\newblock \doi{10.1103/PhysRevD.90.084009}.
\newblock URL \url{https://link.aps.org/doi/10.1103/PhysRevD.90.084009}.

\bibitem[Konoplya et~al.(2016)Konoplya, Rezzolla, and Zhidenko]{PhysRevD.93.064015}
Roman Konoplya, Luciano Rezzolla, and Alexander Zhidenko.
\newblock General parametrization of axisymmetric black holes in metric theories of gravity.
\newblock \emph{Phys. Rev. D}, 93:\penalty0 064015, Mar 2016.
\newblock \doi{10.1103/PhysRevD.93.064015}.
\newblock URL \url{https://link.aps.org/doi/10.1103/PhysRevD.93.064015}.

\bibitem[Randall and Sundrum(1999)]{PhysRevLett.83.4690}
Lisa Randall and Raman Sundrum.
\newblock An alternative to compactification.
\newblock \emph{Phys. Rev. Lett}, 83:\penalty0 4690--4693, Dec 1999.
\newblock \doi{10.1103/PhysRevLett.83.4690}.
\newblock URL \url{https://link.aps.org/doi/10.1103/PhysRevLett.83.4690}.

\bibitem[Yunes and Stein(2011)]{PhysRevD.83.104002}
Nicol\'as Yunes and Leo~C. Stein.
\newblock Nonspinning black holes in alternative theories of gravity.
\newblock \emph{Phys. Rev. D}, 83:\penalty0 104002, May 2011.
\newblock \doi{10.1103/PhysRevD.83.104002}.
\newblock URL \url{https://link.aps.org/doi/10.1103/PhysRevD.83.104002}.

\bibitem[Yunes and Pretorius(2009)]{PhysRevD.79.084043}
Nicol\'as Yunes and Frans Pretorius.
\newblock Dynamical {Chern-Simons} modified gravity: Spinning black holes in the slow-rotation approximation.
\newblock \emph{Phys. Rev. D}, 79:\penalty0 084043, Apr 2009.
\newblock \doi{10.1103/PhysRevD.79.084043}.
\newblock URL \url{https://link.aps.org/doi/10.1103/PhysRevD.79.084043}.

\bibitem[{Capozziello} and {Fang}(2002)]{2002IJMPD..11..483C}
Salvatore {Capozziello} and L.~Z. {Fang}.
\newblock Curvature quintessence.
\newblock \emph{Int. J. Mod. Phys. D}, 11\penalty0 (4):\penalty0 483--491, January 2002.
\newblock \doi{10.1142/S0218271802002025}.
\newblock URL \url{https://ui.adsabs.harvard.edu/abs/2002IJMPD..11..483C}.

\bibitem[Nojiri and Odintsov(2003)]{PhysRevD.68.123512}
Shin'ichi Nojiri and Sergei~D. Odintsov.
\newblock Modified gravity with negative and positive powers of curvature: Unification of inflation and cosmic acceleration.
\newblock \emph{Phys. Rev. D}, 68:\penalty0 123512, Dec 2003.
\newblock \doi{10.1103/PhysRevD.68.123512}.
\newblock URL \url{https://link.aps.org/doi/10.1103/PhysRevD.68.123512}.

\bibitem[{Sotiriou} and {Faraoni}(2010)]{2010RvMP...82..451S}
Thomas~P. {Sotiriou} and Valerio {Faraoni}.
\newblock $f({R})$ theories of gravity.
\newblock \emph{Rev. Mod. Phys.}, 82\penalty0 (1):\penalty0 451--497, January 2010.
\newblock \doi{10.1103/RevModPhys.82.451}.
\newblock URL \url{https://ui.adsabs.harvard.edu/abs/2010RvMP...82..451S}.

\bibitem[Capozziello et~al.(2007)Capozziello, Cardone, and Troisi]{10.1111/j.1365-2966.2007.11401.x}
S.~Capozziello, V.~F Cardone, and A.~Troisi.
\newblock Low surface brightness galaxy rotation curves in the low energy limit of {$R^n$} gravity: no need for dark matter?
\newblock \emph{Mon. Not. R. Astron. Soc.}, 375\penalty0 (4):\penalty0 1423--1440, 02 2007.
\newblock ISSN 0035-8711.
\newblock \doi{10.1111/j.1365-2966.2007.11401.x}.
\newblock URL \url{https://doi.org/10.1111/j.1365-2966.2007.11401.x}.

\bibitem[{Nojiri} and {Odintsov}(2011)]{2011PhR...505...59N}
Shin'Ichi {Nojiri} and Sergei~D. {Odintsov}.
\newblock {Unified cosmic history in modified gravity: From F(R) theory to Lorentz non-invariant models}.
\newblock \emph{Phys. Rep}, 505\penalty0 (2):\penalty0 59--144, August 2011.
\newblock \doi{10.1016/j.physrep.2011.04.001}.
\newblock URL \url{https://ui.adsabs.harvard.edu/abs/2011PhR...505...59N}.

\bibitem[{Nojiri} et~al.(2017){Nojiri}, {Odintsov}, and {Oikonomou}]{2017PhR...692....1N}
S.~{Nojiri}, S.~D. {Odintsov}, and V.~K. {Oikonomou}.
\newblock {Modified gravity theories on a nutshell: Inflation, bounce and late-time evolution}.
\newblock \emph{Phys. Rep}, 692:\penalty0 1--104, June 2017.
\newblock \doi{10.1016/j.physrep.2017.06.001}.
\newblock URL \url{https://ui.adsabs.harvard.edu/abs/2017PhR...692....1N}.

\bibitem[{Nojiri} and {Odintsov}(2003)]{2003PhLB..576....5N}
Shin'ichi {Nojiri} and Sergei~D. {Odintsov}.
\newblock Where new gravitational physics comes from: M-theory?
\newblock \emph{Phys. Lett. B}, 576\penalty0 (1-2):\penalty0 5--11, December 2003.
\newblock \doi{10.1016/j.physletb.2003.09.091}.
\newblock URL \url{https://ui.adsabs.harvard.edu/abs/2003PhLB..576....5N}.

\bibitem[Nojiri et~al.(2006)Nojiri, Odintsov, and Sami]{PhysRevD.74.046004}
Shin'ichi Nojiri, Sergei~D. Odintsov, and M.~Sami.
\newblock Dark energy cosmology from higher-order, string-inspired gravity, and its reconstruction.
\newblock \emph{Phys. Rev. D}, 74:\penalty0 046004, Aug 2006.
\newblock \doi{10.1103/PhysRevD.74.046004}.
\newblock URL \url{https://link.aps.org/doi/10.1103/PhysRevD.74.046004}.

\bibitem[{Kalita}(2018)]{2018ApJ...855...70K}
Sanjeev {Kalita}.
\newblock {Gravitational Theories near the Galactic C}enter.
\newblock \emph{Astrophys. J.}, 855\penalty0 (1):\penalty0 70, March 2018.
\newblock \doi{10.3847/1538-4357/aaadbb}.
\newblock URL \url{https://ui.adsabs.harvard.edu/abs/2018ApJ...855...70K}.

\bibitem[{Kalita}(2020)]{2020ApJ...893...31K}
Sanjeev {Kalita}.
\newblock {The Galactic Center Black Hole, Sgr A*, as a Probe of New Gravitational Physics with the Scalaron Fifth F}orce.
\newblock \emph{Astrophys. J.}, 893\penalty0 (1):\penalty0 31, April 2020.
\newblock \doi{10.3847/1538-4357/ab7af7}.
\newblock URL \url{https://ui.adsabs.harvard.edu/abs/2020ApJ...893...31K}.

\bibitem[{de Martino} et~al.(2020){de Martino}, {Chakrabarty}, {Cesare}, {Gallo}, {Ostorero}, and {Diaferio}]{2020Univ....6..107D}
Ivan {de Martino}, Sankha~S. {Chakrabarty}, Valentina {Cesare}, Arianna {Gallo}, Luisa {Ostorero}, and Antonaldo {Diaferio}.
\newblock {Dark Matters on the Scale of Galaxies}.
\newblock \emph{Universe}, 6\penalty0 (8):\penalty0 107, August 2020.
\newblock \doi{10.3390/universe6080107}.
\newblock URL \url{https://ui.adsabs.harvard.edu/abs/2020Univ....6..107D}.

\bibitem[{Abuter} et~al.(2020)]{2020A&A...636L...5G}
R.~{Abuter} et~al.
\newblock \emph{Astron. Astrophys.}, 636:\penalty0 L5, April 2020.
\newblock \doi{10.1051/0004-6361/202037813}.
\newblock URL \url{https://ui.adsabs.harvard.edu/abs/2020A&A...636L...5G}.

\bibitem[Hees et~al.(2017)Hees, Do, Ghez, Martinez, Naoz, Becklin, Boehle, Chappell, Chu, Dehghanfar, Kosmo, Lu, Matthews, Morris, Sakai, Sch\"odel, and Witzel]{PhysRevLett.118.211101}
A.~Hees, T.~Do, A.~M. Ghez, G.~D. Martinez, S.~Naoz, E.~E. Becklin, A.~Boehle, S.~Chappell, D.~Chu, A.~Dehghanfar, K.~Kosmo, J.~R. Lu, K.~Matthews, M.~R. Morris, S.~Sakai, R.~Sch\"odel, and G.~Witzel.
\newblock Testing general relativity with stellar orbits around the supermassive black hole in our galactic center.
\newblock \emph{Phys. Rev. Lett}, 118:\penalty0 211101, May 2017.
\newblock \doi{10.1103/PhysRevLett.118.211101}.
\newblock URL \url{https://link.aps.org/doi/10.1103/PhysRevLett.118.211101}.

\bibitem[{Abd El Dayem} et~al.(2025)]{2025A&A...698L..15G}
K.~{Abd El Dayem} et~al.
\newblock {Exploring the presence of a fifth force at the Galactic Center}.
\newblock \emph{Astron. Astrophys.}, 698:\penalty0 L15, June 2025.
\newblock \doi{10.1051/0004-6361/202554676}.
\newblock URL \url{https://ui.adsabs.harvard.edu/abs/2025A&A...698L..15G}.

\bibitem[{Paul} et~al.(2024){Paul}, {Bhattacharjee}, and {Kalita}]{2024ApJ...964..127P}
Debojit {Paul}, Pranjali {Bhattacharjee}, and Sanjeev {Kalita}.
\newblock {Kerr-scalaron Metric and Astronomical Consequences near the Galactic Center Black H}ole.
\newblock \emph{Astrophys. J.}, 964\penalty0 (2):\penalty0 127, April 2024.
\newblock \doi{10.3847/1538-4357/ad24f0}.
\newblock URL \url{https://ui.adsabs.harvard.edu/abs/2024ApJ...964..127P}.

\bibitem[{Johannsen} and {Psaltis}(2010)]{2010ApJ...718..446J}
Tim {Johannsen} and Dimitrios {Psaltis}.
\newblock Testing the no-hair theorem with observations in the {Electromagnetic Spectrum. II. Black Hole I}mages.
\newblock \emph{Astrophys. J.}, 718\penalty0 (1):\penalty0 446--454, July 2010.
\newblock \doi{10.1088/0004-637X/718/1/446}.
\newblock URL \url{https://ui.adsabs.harvard.edu/abs/2010ApJ...718..446J}.

\bibitem[{Glampedakis} and {Babak}(2006)]{2006CQGra..23.4167G}
Kostas {Glampedakis} and Stanislav {Babak}.
\newblock {Mapping spacetimes with LISA: inspiral of a test body in a 'quasi-Kerr' field}.
\newblock \emph{Class. Quantum Gravity}, 23\penalty0 (12):\penalty0 4167--4188, June 2006.
\newblock \doi{10.1088/0264-9381/23/12/013}.
\newblock URL \url{https://ui.adsabs.harvard.edu/abs/2006CQGra..23.4167G}.

\bibitem[{Medeiros} et~al.(2020){Medeiros}, {Psaltis}, and {{\"O}zel}]{2020ApJ...896....7M}
Lia {Medeiros}, Dimitrios {Psaltis}, and Feryal {{\"O}zel}.
\newblock \emph{Astrophys. J.}, 896\penalty0 (1):\penalty0 7, June 2020.
\newblock \doi{10.3847/1538-4357/ab8bd1}.
\newblock URL \url{https://ui.adsabs.harvard.edu/abs/2020ApJ...896....7M}.

\bibitem[{Vagnozzi} et~al.(2023){Vagnozzi}, {Roy}, {Tsai}, {Visinelli}, {Afrin}, {Allahyari}, {Bambhaniya}, {Dey}, {Ghosh}, {Joshi}, {Jusufi}, {Khodadi}, {Walia}, {{\"O}vg{\"u}n}, and {Bambi}]{2023CQGra..40p5007V}
Sunny {Vagnozzi}, Rittick {Roy}, Yu-Dai {Tsai}, Luca {Visinelli}, Misba {Afrin}, Alireza {Allahyari}, Parth {Bambhaniya}, Dipanjan {Dey}, Sushant~G. {Ghosh}, Pankaj~S. {Joshi}, Kimet {Jusufi}, Mohsen {Khodadi}, Rahul~Kumar {Walia}, Ali {{\"O}vg{\"u}n}, and Cosimo {Bambi}.
\newblock {Horizon-scale tests of gravity theories and fundamental physics from the Event Horizon Telescope image of Sagittarius A (*)}.
\newblock \emph{Class. Quantum Gravity}, 40\penalty0 (16):\penalty0 165007, August 2023.
\newblock \doi{10.1088/1361-6382/acd97b}.
\newblock URL \url{https://ui.adsabs.harvard.edu/abs/2023CQGra..40p5007V}.

\bibitem[{Nojiri} and {Odintsov}(2025)]{2025PDU....4701785N}
Shin'ichi {Nojiri} and S.~D. {Odintsov}.
\newblock {Black holes and their shadows in $f(R)$ gravity}.
\newblock \emph{Phys. Dark Univ.}, 47:\penalty0 101785, February 2025.
\newblock \doi{10.1016/j.dark.2024.101785}.
\newblock URL \url{https://ui.adsabs.harvard.edu/abs/2025PDU....4701785N}.

\bibitem[{Tsai} et~al.(2023){Tsai}, {Wu}, {Vagnozzi}, and {Visinelli}]{2023JCAP...04..031T}
Yu-Dai {Tsai}, Youjia {Wu}, Sunny {Vagnozzi}, and Luca {Visinelli}.
\newblock {Novel constraints on fifth forces and ultralight dark sector with asteroidal data}.
\newblock \emph{JCAP}, 2023\penalty0 (4):\penalty0 031, April 2023.
\newblock \doi{10.1088/1475-7516/2023/04/031}.
\newblock URL \url{https://ui.adsabs.harvard.edu/abs/2023JCAP...04..031T}.

\bibitem[{Tsai} et~al.(2024){Tsai}, {Farnocchia}, {Micheli}, {Vagnozzi}, and {Visinelli}]{2024CmPhy...7..311T}
Yu-Dai {Tsai}, Davide {Farnocchia}, Marco {Micheli}, Sunny {Vagnozzi}, and Luca {Visinelli}.
\newblock {Constraints on fifth forces and ultralight dark matter from OSIRIS-REx target asteroid Bennu}.
\newblock \emph{Communications Physics}, 7\penalty0 (1):\penalty0 311, December 2024.
\newblock \doi{10.1038/s42005-024-01779-3}.
\newblock URL \url{https://ui.adsabs.harvard.edu/abs/2024CmPhy...7..311T}.

\bibitem[{Khodadi} et~al.(2024){Khodadi}, {Vagnozzi}, and {Firouzjaee}]{2024NatSR..1426932K}
Mohsen {Khodadi}, Sunny {Vagnozzi}, and Javad~T. {Firouzjaee}.
\newblock {Event Horizon Telescope observations exclude compact objects in baseline mimetic gravity}.
\newblock \emph{Scientific Reports}, 14\penalty0 (1):\penalty0 26932, November 2024.
\newblock \doi{10.1038/s41598-024-78264-y}.
\newblock URL \url{https://ui.adsabs.harvard.edu/abs/2024NatSR..1426932K}.

\bibitem[{Bambi} et~al.(2019){Bambi}, {Freese}, {Vagnozzi}, and {Visinelli}]{2019PhRvD.100d4057B}
Cosimo {Bambi}, Katherine {Freese}, Sunny {Vagnozzi}, and Luca {Visinelli}.
\newblock {Testing the rotational nature of the supermassive object M87* from the circularity and size of its first image}.
\newblock \emph{Phys. Rev. D}, 100\penalty0 (4):\penalty0 044057, August 2019.
\newblock \doi{10.1103/PhysRevD.100.044057}.
\newblock URL \url{https://ui.adsabs.harvard.edu/abs/2019PhRvD.100d4057B}.

\bibitem[{Paul} et~al.(2023){Paul}, {Kalita}, and {Talukdar}]{2023IJMPD..3250021P}
Debojit {Paul}, Sanjeev {Kalita}, and Abhijit {Talukdar}.
\newblock Unscreening of {f(R)} gravity near the galactic center black hole: Testability through pericenter shift below {S0-2’s} orbit.
\newblock \emph{Int. J. Mod. Phys. D}, 32\penalty0 (4):\penalty0 2350021-91, January 2023.
\newblock \doi{10.1142/S0218271823500219}.
\newblock URL \url{https://ui.adsabs.harvard.edu/abs/2023IJMPD..3250021P}.

\bibitem[{Kalita} and {Bhattacharjee}(2023)]{2023EPJC...83..120K}
Sanjeev {Kalita} and Pranjali {Bhattacharjee}.
\newblock Constraining spacetime metrics within and outside general relativity through the galactic center black hole {(Sgr A*)} shadow.
\newblock \emph{Eur. Phys. J. C}, 83\penalty0 (2):\penalty0 120, February 2023.
\newblock \doi{10.1140/epjc/s10052-023-11226-2}.
\newblock URL \url{https://ui.adsabs.harvard.edu/abs/2023EPJC...83..120K}.

\bibitem[Abbott et~al.(2021)]{PhysRevD.103.122002}
R.~Abbott et~al.
\newblock Tests of general relativity with binary black holes from the second {LIGO-Virgo} gravitational-wave transient catalog.
\newblock \emph{Phys. Rev. D}, 103:\penalty0 122002, Jun 2021.
\newblock \doi{10.1103/PhysRevD.103.122002}.
\newblock URL \url{https://link.aps.org/doi/10.1103/PhysRevD.103.122002}.

\bibitem[{Amorim} et~al.(2019)]{10.1093/mnras/stz2300}
A.~{Amorim} et~al.
\newblock Scalar field effects on the orbit of s2 star.
\newblock \emph{Mon. Not. R. Astron. Soc.}, 489\penalty0 (4):\penalty0 4606--4621, 11 2019.
\newblock ISSN 0035-8711.
\newblock \doi{10.1093/mnras/stz2300}.
\newblock URL \url{https://doi.org/10.1093/mnras/stz2300}.

\bibitem[Arvanitaki et~al.(2010)Arvanitaki, Dimopoulos, Dubovsky, Kaloper, and March-Russell]{PhysRevD.81.123530}
Asimina Arvanitaki, Savas Dimopoulos, Sergei Dubovsky, Nemanja Kaloper, and John March-Russell.
\newblock String axiverse.
\newblock \emph{Phys. Rev. D}, 81:\penalty0 123530, Jun 2010.
\newblock \doi{10.1103/PhysRevD.81.123530}.
\newblock URL \url{https://link.aps.org/doi/10.1103/PhysRevD.81.123530}.

\bibitem[{Paul} and {Kalita}(2025)]{2025PhyS..100f5006P}
Debojit {Paul} and Sanjeev {Kalita}.
\newblock {f(R)} gravity in the solar system and cosmological scalarons.
\newblock \emph{Phys. Scr.}, 100\penalty0 (6):\penalty0 065006, June 2025.
\newblock \doi{10.1088/1402-4896/add225}.
\newblock URL \url{https://ui.adsabs.harvard.edu/abs/2025PhyS..100f5006P}.

\bibitem[Khan and Ren(2020)]{KHAN2020100644}
Saeed~Ullah Khan and Jingli Ren.
\newblock Shadow cast by a rotating charged black hole in quintessential dark energy.
\newblock \emph{Phys. Dark Univ.}, 30:\penalty0 100644, 2020.
\newblock ISSN 2212-6864.
\newblock URL \url{https://www.sciencedirect.com/science/article/pii/S2212686420301941}.

\bibitem[Khan and Ren(2022)]{KHAN2022141}
Saeed~Ullah Khan and Jingli Ren.
\newblock Geodesics and optical properties of a rotating black hole in randall–sundrum brane with a cosmological constant.
\newblock \emph{Chinese Journal of Physics}, 78:\penalty0 141--154, 2022.
\newblock ISSN 0577-9073.
\newblock URL \url{https://www.sciencedirect.com/science/article/pii/S0577907322001691}.

\bibitem[Khan et~al.(2024)Khan, Rayimbaev, Sarikulov, and Abdurakhmonov]{KHAN2024690}
Saeed~Ullah Khan, Javlon Rayimbaev, Furkat Sarikulov, and Ozodbek Abdurakhmonov.
\newblock Optical features of rotating quintessential charged black holes in de-sitter spacetime.
\newblock \emph{Chin. J. Phys.}, 90:\penalty0 690--706, 2024.
\newblock ISSN 0577-9073.
\newblock URL \url{https://www.sciencedirect.com/science/article/pii/S0577907324002302}.

\bibitem[Khan et~al.(2019)Khan, Shahzadi, and Ren]{KHAN2019100331}
Saeed~Ullah Khan, Misbah Shahzadi, and Jingli Ren.
\newblock Particle collisions in ergoregion of braneworld kerr black hole.
\newblock \emph{Phys. Dark Univ.}, 26:\penalty0 100331, 2019.
\newblock ISSN 2212-6864.
\newblock URL \url{https://www.sciencedirect.com/science/article/pii/S2212686419300743}.

\bibitem[{Khan} et~al.(2026){Khan}, {Rayimbaev}, {Turaev}, {Sirajiddin}, {Usanov}, and {Wang}]{2026EPJC...86..207K}
Saeed~Ullah {Khan}, Javlon {Rayimbaev}, Yunus {Turaev}, Otaboyev {Sirajiddin}, Sulton {Usanov}, and Weiwei {Wang}.
\newblock {Energy efficiency and particle dynamics around magnetized black holes with parabolic configuration in STVG}.
\newblock \emph{Eur. Phys. J. C}, 86\penalty0 (3):\penalty0 207, March 2026.
\newblock URL \url{https://ui.adsabs.harvard.edu/abs/2026EPJC...86..207K}.

\bibitem[Khan et~al.(2026)Khan, Zahid, Chen, Turaev, Akhmedov, Usanov, Rayimbaev, and Wang]{KHAN2026711}
Saeed~Ullah Khan, Muhammad Zahid, Zhi-Min Chen, Yunus Turaev, Munisbek Akhmedov, Sulton Usanov, Javlon Rayimbaev, and Weiwei Wang.
\newblock Rotating black-bounce black holes in modified gravity: Photon orbits and shadow.
\newblock \emph{Chin. J. Phys.}, 102:\penalty0 711--727, 2026.
\newblock ISSN 0577-9073.
\newblock \doi{https://doi.org/10.1016/j.cjph.2026.03.023}.
\newblock URL \url{https://www.sciencedirect.com/science/article/pii/S0577907326000870}.

\bibitem[Psaltis(2009)]{DimitriosPsaltis_2009}
Dimitrios Psaltis.
\newblock Two approaches to testing general relativity in the strong-field regime.
\newblock \emph{Journal of Physics: Conference Series}, 189\penalty0 (1):\penalty0 012033, oct 2009.
\newblock \doi{10.1088/1742-6596/189/1/012033}.
\newblock URL \url{https://doi.org/10.1088/1742-6596/189/1/012033}.

\bibitem[{Sakharov}(1968)]{1968SPhD...12.1040S}
A.~D. {Sakharov}.
\newblock {Vacuum Quantum Fluctuations in Curved Space and the Theory of Gravitation}.
\newblock \emph{Soviet Physics Doklady}, 12:\penalty0 1040, May 1968.
\newblock URL \url{https://ui.adsabs.harvard.edu/abs/1968SPhD...12.1040S}.

\bibitem[{Ruzma{\v{i}}kina} and {Ruzma{\v{i}}kin}(1969)]{1969JETP...30..372R}
T.~V. {Ruzma{\v{i}}kina} and A.~A. {Ruzma{\v{i}}kin}.
\newblock {Quadratic Corrections to the Lagrangian Density of the Gravitational Field and the Singularity}.
\newblock \emph{Soviet Journal of Experimental and Theoretical Physics}, 30:\penalty0 372, January 1969.
\newblock URL \url{https://ui.adsabs.harvard.edu/abs/1969JETP...30..372R}.

\bibitem[{Talukdar} et~al.(2024){Talukdar}, {Kalita}, {Das}, and {Lahkar}]{2024JCAP...02..019T}
Abhijit {Talukdar}, Sanjeev {Kalita}, Nirmali {Das}, and Nandita {Lahkar}.
\newblock Constraining primordial black hole masses through {f(R) gravity scalarons in Big Bang N}ucleosynthesis.
\newblock \emph{JCAP}, 2024\penalty0 (2):\penalty0 019, February 2024.
\newblock \doi{10.1088/1475-7516/2024/02/019}.
\newblock URL \url{https://ui.adsabs.harvard.edu/abs/2024JCAP...02..019T}.

\bibitem[{Chandrasekhar}(1983)]{1983mtbh.book.....C}
S.~{Chandrasekhar}.
\newblock \emph{{The mathematical theory of black holes}}.
\newblock Clarendon Press, Oxford, 1983.
\newblock URL \url{https://ui.adsabs.harvard.edu/abs/1983mtbh.book.....C}.

\bibitem[{Luminet}(1979)]{1979A&A....75..228L}
J.-P. {Luminet}.
\newblock Image of a spherical black hole with thin accretion disk.
\newblock \emph{Astron. Astrophys.}, 75:\penalty0 228--235, May 1979.
\newblock URL \url{https://ui.adsabs.harvard.edu/abs/1979A&A....75..228L}.

\bibitem[{V{\'a}zquez} and {Esteban}(2004)]{2004NCimB.119..489V}
S.~E. {V{\'a}zquez} and E.~P. {Esteban}.
\newblock Strong-field gravitational lensing by a kerr black hole.
\newblock \emph{Nuovo Cimento B Serie}, 119\penalty0 (5):\penalty0 489, May 2004.
\newblock \doi{10.1393/ncb/i2004-10121-y}.
\newblock URL \url{https://ui.adsabs.harvard.edu/abs/2004NCimB.119..489V}.

\bibitem[Bambi et~al.(2019)Bambi, Freese, Vagnozzi, and Visinelli]{PhysRevD.100.044057}
Cosimo Bambi, Katherine Freese, Sunny Vagnozzi, and Luca Visinelli.
\newblock Testing the rotational nature of the supermassive object m87* from the circularity and size of its first image.
\newblock \emph{Phys. Rev. D}, 100:\penalty0 044057, Aug 2019.
\newblock \doi{10.1103/PhysRevD.100.044057}.
\newblock URL \url{https://link.aps.org/doi/10.1103/PhysRevD.100.044057}.

\bibitem[{Shaikh}(2023)]{2023MNRAS.523..375S}
Rajibul {Shaikh}.
\newblock {Testing black hole mimickers with the Event Horizon Telescope image of Sagittarius A*}.
\newblock \emph{Mon. Not. R. Astron. Soc.}, 523\penalty0 (1):\penalty0 375--384, July 2023.
\newblock \doi{10.1093/mnras/stad1383}.
\newblock URL \url{https://ui.adsabs.harvard.edu/abs/2023MNRAS.523..375S}.

\bibitem[Gralla et~al.(2019)Gralla, Holz, and Wald]{PhysRevD.100.024018}
Samuel~E. Gralla, Daniel~E. Holz, and Robert~M. Wald.
\newblock Black hole shadows, photon rings, and lensing rings.
\newblock \emph{Phys. Rev. D}, 100:\penalty0 024018, Jul 2019.
\newblock \doi{10.1103/PhysRevD.100.024018}.
\newblock URL \url{https://link.aps.org/doi/10.1103/PhysRevD.100.024018}.

\bibitem[Hobson et~al.(2006)Hobson, Efstathiou, and Lasenby]{Hobson:2006se}
M.~P. Hobson, G.~P. Efstathiou, and A.~N. Lasenby.
\newblock \emph{{General relativity: An introduction for physicists}}.
\newblock Cambridge University Press, 2006.

\bibitem[Israel(1967)]{PhysRev.164.1776}
Werner Israel.
\newblock Event horizons in static vacuum space-times.
\newblock \emph{Phys. Rev.}, 164:\penalty0 1776--1779, Dec 1967.
\newblock \doi{10.1103/PhysRev.164.1776}.
\newblock URL \url{https://link.aps.org/doi/10.1103/PhysRev.164.1776}.

\bibitem[Carter(1971)]{PhysRevLett.26.331}
B.~Carter.
\newblock Axisymmetric black hole has only two degrees of freedom.
\newblock \emph{Phys. Rev. Lett.}, 26:\penalty0 331--333, Feb 1971.
\newblock \doi{10.1103/PhysRevLett.26.331}.
\newblock URL \url{https://link.aps.org/doi/10.1103/PhysRevLett.26.331}.

\bibitem[{Hawking}(1972)]{1972CMaPh..25..152H}
S.~W. {Hawking}.
\newblock {Black holes in general relativity}.
\newblock \emph{Communications in Mathematical Physics}, 25\penalty0 (2):\penalty0 152--166, June 1972.
\newblock \doi{10.1007/BF01877517}.
\newblock URL \url{https://ui.adsabs.harvard.edu/abs/1972CMaPh..25..152H}.

\bibitem[Thorne(1980)]{RevModPhys.52.299}
Kip~S. Thorne.
\newblock Multipole expansions of gravitational radiation.
\newblock \emph{Rev. Mod. Phys.}, 52:\penalty0 299--339, Apr 1980.
\newblock \doi{10.1103/RevModPhys.52.299}.
\newblock URL \url{https://link.aps.org/doi/10.1103/RevModPhys.52.299}.

\bibitem[{Baker} et~al.(2015){Baker}, {Psaltis}, and {Skordis}]{2015ApJ...802...63B}
Tessa {Baker}, Dimitrios {Psaltis}, and Constantinos {Skordis}.
\newblock {Linking Tests of Gravity on All Scales: from the Strong-field Regime to Cosmology}.
\newblock \emph{Astrophys. J.}, 802\penalty0 (1):\penalty0 63, March 2015.
\newblock \doi{10.1088/0004-637X/802/1/63}.
\newblock URL \url{https://ui.adsabs.harvard.edu/abs/2015ApJ...802...63B}.

\bibitem[{Jain} et~al.(2013){Jain}, {Vikram}, and {Sakstein}]{2013ApJ...779...39J}
Bhuvnesh {Jain}, Vinu {Vikram}, and Jeremy {Sakstein}.
\newblock {Astrophysical Tests of Modified Gravity: Constraints from Distance Indicators in the Nearby Universe}.
\newblock \emph{Astrophys. J.}, 779\penalty0 (1):\penalty0 39, December 2013.
\newblock \doi{10.1088/0004-637X/779/1/39}.
\newblock URL \url{https://ui.adsabs.harvard.edu/abs/2013ApJ...779...39J}.

\bibitem[Khoury and Weltman(2004)]{PhysRevLett.93.171104}
Justin Khoury and Amanda Weltman.
\newblock Chameleon fields: Awaiting surprises for tests of gravity in space.
\newblock \emph{Phys. Rev. Lett}, 93:\penalty0 171104, Oct 2004.
\newblock \doi{10.1103/PhysRevLett.93.171104}.
\newblock URL \url{https://link.aps.org/doi/10.1103/PhysRevLett.93.171104}.

\bibitem[Hu and Sawicki(2007)]{PhysRevD.76.064004}
Wayne Hu and Ignacy Sawicki.
\newblock Models of $f(r)$ cosmic acceleration that evade solar system tests.
\newblock \emph{Phys. Rev. D}, 76:\penalty0 064004, Sep 2007.
\newblock \doi{10.1103/PhysRevD.76.064004}.
\newblock URL \url{https://link.aps.org/doi/10.1103/PhysRevD.76.064004}.

\bibitem[Sharma et~al.(2024)Sharma, Harikumar, Grespan, Biesiada, and Verma]{SHARMA2024139093}
Vipin~Kumar Sharma, Sreekanth Harikumar, Margherita Grespan, Marek Biesiada, and Murli~Manohar Verma.
\newblock Probing massive gravitons in {f(R)} with lensed gravitational waves.
\newblock \emph{Phys. Lett. B}, 859:\penalty0 139093, 2024.
\newblock ISSN 0370-2693.
\newblock \doi{https://doi.org/10.1016/j.physletb.2024.139093}.
\newblock URL \url{https://www.sciencedirect.com/science/article/pii/S0370269324006518}.

\bibitem[Lemos et~al.(2024)Lemos, Campos, and Brito]{PhysRevD.110.064079}
A.~S. Lemos, J.~A.~V. Campos, and F.~A. Brito.
\newblock Hunting for extra dimensions in black hole shadows.
\newblock \emph{Phys. Rev. D}, 110:\penalty0 064079, Sep 2024.
\newblock \doi{10.1103/PhysRevD.110.064079}.
\newblock URL \url{https://link.aps.org/doi/10.1103/PhysRevD.110.064079}.

\end{thebibliography}






\end{document}